%% file: Drton_Maathuis_Graphical_Models_Arxiv.tex
\documentclass{amsart}

\usepackage{amsmath,amsthm,amssymb}
\usepackage{natbib}
\usepackage[left=1.4in,right=1.4in,top=1.4in,bottom=1.4in]{geometry}

\usepackage{latexsym,graphicx,keyval}

\usepackage{caption,subfig}
\usepackage{tikz}
\usetikzlibrary{shapes,arrows,positioning}

\input{macros}
\numberwithin{equation}{section}

\title{Structure Learning in Graphical Modeling}

\author{Mathias Drton}
\address{Department of Statistics, University of Washington,
  Seattle, U.S.A., 98195}
\email{md5@uw.edu}
\author{Marloes H.\ Maathuis}
\address{Seminar f\"{u}r Statistik, ETH Z\"{u}rich, Z\"{u}rich,
  Switzerland, 8092}
\email{maathuis@stat.math.ethz.ch}

\begin{document}
%Abstract
\begin{abstract}
  A graphical model is a statistical model that is associated to a
  graph whose nodes correspond to variables of interest.  The edges of
  the graph reflect allowed conditional dependencies among
  the variables.  Graphical models admit computationally convenient
  factorization properties and have long been a valuable tool for
  tractable modeling of multivariate distributions.
  More recently,
  applications such as reconstructing gene regulatory
  networks from gene expression data have driven major advances in
  structure learning, that is, estimating the graph underlying a
  model.
  We review some of these advances and discuss methods such as
  the graphical lasso and neighborhood selection for undirected
  graphical models (or Markov random fields), and the PC algorithm and
  score-based search methods for directed graphical models (or
  Bayesian networks).  We further review extensions that account for
  effects of latent variables and heterogeneous data sources.
\end{abstract}

\keywords{Bayesian network, graphical model, Markov random field, model
selection, multivariate statistics, network reconstruction}

\maketitle

\section{Introduction}
\label{sec:introduction}

This article gives an overview of commonly used techniques for
structure learning in graphical modeling.  Structure learning is a
model selection problem in which one estimates a graph that summarizes
the dependence structure in a given data set.

\subsection{What is a Graphical Model?}
\label{sec:what-graphical-model}

A graphical model captures stochastic dependencies among a collection of
random variables $X_v$, $v\in V$ \citep{Lauritzen96}.  Each
model is associated to a graph $G=(V,E)$, where the vertex set $V$ indexes
the variables and the edge set $E$ imposes a set of conditional
independencies. Specifically, $X_v$
and $X_w$ are required to be conditionally independent given
$X_C:=(X_u: u\in C)$, denoted by
$X_v\indep X_w \,|\, X_C$, if every path between nodes $v$ and $w$ in
$G$ is suitably blocked by the nodes in $C$.
% \begin{marginnote}[]
%   \entry{Graphical models}{Conditional independence models for random vectors}
% \end{marginnote}

Markov chains constitute a familiar example of graphical models.  We
first demonstrate this in the context of undirected graphs, for which
the edges are unordered pairs $\{v,w\}$ for distinct $v,w\in V$.
We also write
$v-w\in E$.
In an undirected graph, a path is \emph{blocked} by $C$ if
it contains a node in $C$.

\begin{figure} %[h]
  \centering
  \subfloat[An undirected graph.]{
    \label{fig:markov-chains-a}
    \begin{tikzpicture}[auto,
      main node/.style={circle,inner
        sep=2pt,fill=gray!20,draw,font=\sffamily}]

      \node[main node,rounded corners] (1) {1}; \node[main
      node,rounded corners] (2) [right =0.8cm of 1] {2}; \node[main
      node,rounded corners] (3) [right =0.8cm of 2] {3}; \node[main
      node,rounded corners] (4) [right =0.8cm of 3] {4}; \node[main
      node,rounded corners] (5) [right =0.8cm of 4] {5};
      % just for vertical spacing of caption
      \node[main
      node,rounded corners,draw=none,fill=none] (6) [below =.0cm of 4] {};

      \path[every
      node/.style={font=\sffamily\small}] (1) edge node {} (2) (2)
      edge node {} (3) (3) edge node {} (4) (4) edge node {} (5);
    \end{tikzpicture}
    }
    \hfill
    %\\[.25cm]
    \captionsetup[subfigure]{width=1.85in}
    \subfloat[A directed acyclic graph (DAG).] 
    {
    \label{fig:markov-chains-b}
    \begin{tikzpicture}[->,>=stealth', shorten >=1pt, auto,
      main node/.style={circle,inner
        sep=2pt,fill=gray!20,draw,font=\sffamily}]

      \node[main node,rounded corners] (1) {1}; \node[main node,rounded
      corners] (2) [right =0.8cm of 1] {2}; \node[main node,rounded
      corners] (3) [right =0.8cm of 2] {3}; \node[main node,rounded
      corners] (4) [right =0.8cm of 3] {4}; \node[main node,rounded
      corners] (5) [right =0.8cm of 4] {5};
            % just for vertical spacing of caption
      \node[main
      node,rounded corners,draw=none,fill=none] (6) [below =.0cm of 4]
      {};

      \path[every
      node/.style={font=\sffamily\small}] (1) edge node {} (2) (2) edge
      node {} (3) (3) edge node {} (4) (4) edge node {} (5);
    \end{tikzpicture}
  }
  \caption{Two graphs that both induce a Markov chain.}
  \label{fig:markov-chains}
\end{figure}
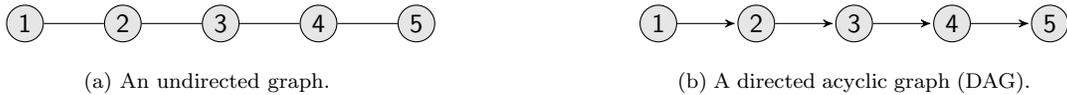

\begin{example}
  Suppose that
  $(X_1,\dots,X_5)$ belongs to the graphical model for
  the undirected graph in Figure \ref{fig:markov-chains-a}. %
  Then $X_v\indep X_w \,|\, X_C$ whenever $C$ contains a node $c$ on
  the unique path between $v$ and $w$, so $v<c<w$ or $w<c<v$.  For
  example, $X_2\indep X_4\,|\, (X_1,X_3)$.  Thinking of the indices as
  time, the past and the future are conditionally independent given
  the present. We recognize that $X_1,\dots,X_5$ form a Markov chain.
  The generalization to a Markov chain of arbitrary length is
  obvious.
\end{example}
% \begin{marginnote}[]
%   \entry{Graphical models}{Generalizations of Markov chains}
% \end{marginnote}

Let $\nb_G(v)=\{w \in V:\{w,v\}\in E\}$ be the neighbors of node $v$ in an undirected graph $G=(V,E)$.  As
detailed in Section~\ref{sec:basic-conc-graph}, a typically equivalent
interpretation of the graph is that
each $X_v$ is conditionally independent of its non-neighbors
$X_{V\setminus (\nb_G(v)\cup\{v\})}$ given its neighbors $X_{\nb_G(v)}$.
A mean squared error optimal
prediction of $X_v$ can thus be made from its neighbors $X_{\nb_G(v)}$
alone.
% \begin{marginnote}[]
%   \entry{Undirected
%     graphs}{Optimal prediction from neighboring variables}
% \end{marginnote}

Graphical models
can also
be built from directed graphs $G=(V,E)$.  The edge set $E$ then comprises
ordered pairs $(v,w)$ that represent an edge pointing from $v$ to $w$.
We also write $v\to w\in E$.
Adopting language for family trees,
let $\pa_G(v)=\{w\in V:(w,v)\in E\}$ be the parents of node $v$ in $G$,
let $\de_G(v) =\{w\in V: w=v \,\, \text{or} \,\, v\to\dots \to w \,\,\text{in}\,\,G\}$ be the descendants of
$v$ in $G$, and let $\nd_G(v) = V \setminus \de_G(v)$ be the non-descendants of $v$ in $G$.
In a directed acyclic graph, we then
require that each variable $X_v$ is conditionally independent of its
non-descendants $X_{\nd_G(v)\setminus\pa_G(v)}$ given its parents
$X_{\pa_G(v)}$.  Such independencies arise when each variable $X_v$ is a
stochastic function of $X_{\pa_G(v)}$.  This is
the starting point for a connection to causal modeling
\citep{Pearl09,SpirtesEtAl00}.
The notion of blocking a path in a directed graph is different and more subtle than blocking in an undirected graph.
We give the details in
Section~\ref{sec:basic-conc-graph}.
% \begin{marginnote}[]
%   \entry{Directed graphs}{Can capture cause-effect relationships}
% \end{marginnote}

\begin{example}
  Suppose that
  $(X_1,\dots,X_5)$ belongs to the graphical model for the
  directed acyclic graph $G$ in Figure~\ref{fig:markov-chains-b}.
  Since $\pa_G(v)=\{v-1\}$ for $v=2,\dots,5$, the graph encodes
  $X_v\indep (X_1,\dots,X_{v-2})\,|\, X_{v-1}$ for $v=3,4,5$.
  This is precisely the standard Markov property of a Markov chain.
  Again, this example is easily generalized to a chain of
  arbitrary length.
\end{example}

Graphical models can also be defined in terms of density
factorizations.  Indeed, much of the popularity of graphical models is
due to
the fact
that factorizations allow efficient storage and computation
with high-dimensional joint distributions
\citep{MAL-001}.
To explain, suppose for simplicity
that $X_1,X_2,\dots,X_m$ are binary
and form a Markov chain, in which case the joint
probability factorizes as
% \begin{marginnote}[]
%   \entry{Density factorization}{Efficient storage and computation}
% \end{marginnote}
\begin{equation}
  \label{eq:intro-dag-factorization}
  \Pr(X_1=x_1,\dots,X_m=x_m)
  \;=\; \Pr(X_1=x_1) \prod_{v=2}^m
    \Pr(X_v=x_v\,|\,X_{v-1}=x_{v-1}).
\end{equation}
The $2^m-1$ dimensional joint distribution is thus determined by
merely $2m-1$ parameters.  In reference to the directed graph $G$ in
Figure~\ref{fig:markov-chains-b}, the product
in~(\ref{eq:intro-dag-factorization}) is over the conditional
probabilities $\Pr(X_v=x_v\,|\, X_{\pa_G(v)}=x_{\pa_G(v)})$, or simply
$\Pr(X_v=x_v)$ when $\pa_G(v)=\emptyset$.
For the undirected graph in Figure~\ref{fig:markov-chains-a}, the
right-hand side of~(\ref{eq:intro-dag-factorization}) may be viewed as
a product of $m-1$ functions whose arguments are the pairs
$(x_v,x_{v+1})$ that correspond to the %$m-1$
edges of the graph.
Section~\ref{sec:basic-conc-graph} gives
generalizations of these
observations.

\subsection{Application:  Reconstruction of Gene Regulatory Networks}

The last decade has seen great advances in structure learning, with
new methods being developed and older methods being viewed in new
light.  These developments have largely been driven by problems in
biology, such as inferring a network of regulatory relationships among
genes from data on their expression levels
\citep{Friedman:2004}.

\begin{figure}
  \centering
  \includegraphics[scale=0.4]{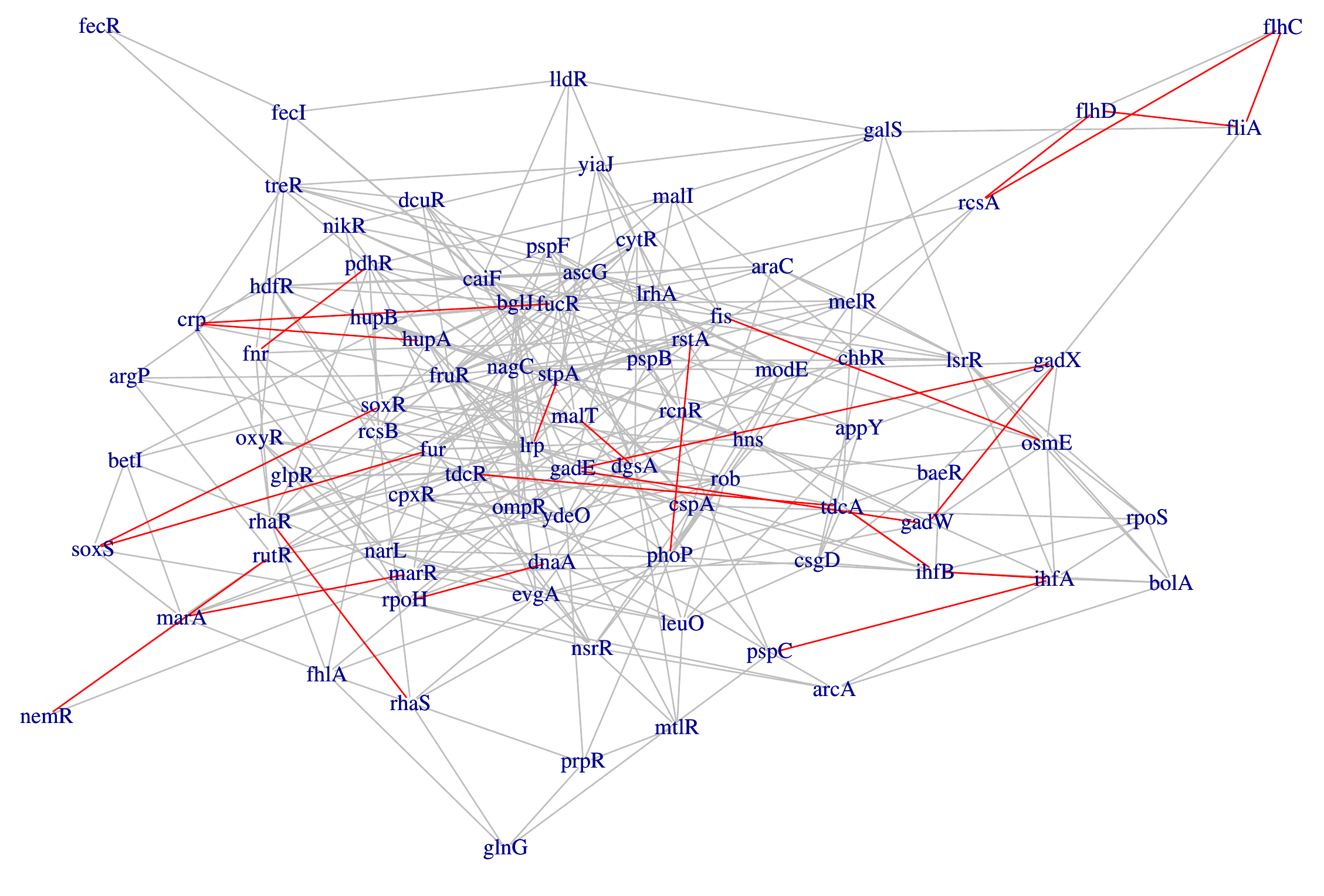}
  \smallskip
  \caption{Estimated conditional independence graph in a Gaussian copula model
    for data on the expression of 87 transcription factors in
    \emph{E.~coli}.}
  \label{fig:dream5}
\end{figure}

\begin{example}\label{ex:Marbach}
  Reporting on a prediction challenge, \cite{wisdom} provide data on
  gene expression in
  \emph{E.~coli}. We restrict attention to the $|V|=87$
  genes that form the only large connected component in the network of
  known interactions among transcription factors (`known' prior to the
  challenge).  Ignoring
  heterogeneity across the $n=804$ samples, we apply
  neighborhood selection in a Gaussian copula model (see
  Sections~\ref{sec:neighborhood-selection} and
  \ref{sec:semiparametric}), under the defaults of the software of
  \cite{MR2930633}.  The estimated undirected conditional independence
  graph (see Section~\ref{sec:undir-graph-models}) has 352 edges.  It
  is plotted in Figure~\ref{fig:dream5}, where each one of the 24 red
  edges corresponds to one of the 124 pairs of transcription factors
  that are known to interact.  While the majority of known
  interactions fails to be part of the estimate, some signal is being
  detected.  The probability that a random selection of 352 edges
  would comprise at least 24 of the 124 known interactions is about
  $1.5\times 10^{-4}$.
\end{example}
% \begin{marginnote}[]
%   \entry{Gene regulatory networks}{Much recent progress in structure
%     learning has been driven by applications to gene expression}
% \end{marginnote}

Example \ref{ex:Marbach}
involves a
number of pre-selected
genes and is thus of moderate dimensionality.
Modern
experiments often yield far higher-dimensional data,
presenting a
statistical challenge that is behind much of the recent fascination
with graphical models.

\subsection{Outline of the Review}

This review begins with more background on graphical models in
Section~\ref{sec:basic-conc-graph}.
Sections~\ref{sec:learn-undir-graph} and \ref{sec:learn-direct-graph} discuss
structure learning
for undirected and directed graphical models.
A brief treatment of issues arising from latent variables and
heterogeneous data sources is given in
Sections~\ref{sec:latent-variables} and~\ref{sec:heterogeneous-data}.  We end with a discussion in
Section~\ref{sec:discussion}.

While our review conveys some of the main ideas in structure learning,
several interesting topics are beyond the scope of our paper.  For
instance, we do not cover Bayesian inference, even though there is an
active Bayesian community whose recent work tackles problems such as
heterogeneous data \citep{MR3338494,MR3447093}, latent variables
\citep{MR2520804}, posterior convergence
rates \citep{MR3321485}, robustness \citep{MR3256052}, sampling Markov
equivalence classes \citep{MR3127848}, and context-specific
independence \citep{MR3293960}.
We also do not treat dynamic graphical models for multivariate time
series,
but a discussion of these models can be found in the related paper by \cite{didelez}.
Another area omitted in this review is active learning;
see, e.g.,~\cite{ActiveVats}, \cite{StatnikovEtAl15}, and
\cite{ActiveDasarthy}.

Finally, the literature on structure learning is vast and our
references are necessarily selective.  To limit the number of citations,
we sometimes only cite a recent paper, trusting that readers will
follow the trail of literature to identify earlier work on the
topic.

%%%%%%%%%%%%%%%%%%%%%%%%%%%%%%%%%%%%

\section{Basic Concepts in Graphical Modeling}
\label{sec:basic-conc-graph}

In this section, we review some
essential concepts for
undirected and directed graphical models
\citep[see, e.g.,][]{Lauritzen96,MR3183760,Pearl09}.

\subsection{Undirected Graphical Models}
\label{sec:undir-graph-models}

Let $X=(X_v:v\in V)$ be a random vector indexed by the vertices of an
undirected graph $G=(V,E)$.  Then $X$
satisfies the
\emph{pairwise Markov property} with respect to $G$ if
\begin{equation}
  \label{eq:ug:pairwiseMP}
  X_v\indep X_w\mid X_{V\setminus\{v,w\}}
\end{equation}
whenever $\{v,w\}\notin E$.
Moreover, $X$ satisfies the \emph{local
Markov property} with respect to $G$ if
\begin{equation}
  \label{eq:ug:localMP}
  X_v\indep X_{V\setminus(\nb_G(v)\cup\{v\})} \mid X_{\nb_G(v)}
\end{equation}
for every $v\in V$, where we recall that
$\nb_G(v)=\{w\in V: \{w,v\}\in E\}$
is the set of neighbors of $v$.
Finally, $X$ satisfies the \emph{global Markov property} with respect to $G$
if
  $X_A\indep X_B \mid X_C$
for all triples of pairwise disjoint subsets $A,B,C\subset V$ such
that $C$ \emph{separates} $A$ and $B$ in $G$, i.e.,
such that every path between a node in $A$
and a node in $B$ contains a node in $C$.

It is easy to see that the global Markov property implies the
pairwise and local properties.
It can also be shown that the local Markov
property implies the pairwise Markov property.  While not true in
general, the three Markov properties are equivalent when $X$ satisfies
the intersection axiom for conditional independence
\citep{Lauritzen96}.
This equivalence is true in particular when the
$X_v$ are discrete with positive joint probabilities,
or when $X$ has a positive and continuous density with respect to
Lebesgue measure.

% \begin{marginnote}[]
%    \entry{Markov properties for undirected graphs}{In general: global $\Rightarrow$ local
%      $\Rightarrow$ pairwise
%    \smallskip}
%      {
%      Under intersection axiom: global $\Leftrightarrow$ local $\Leftrightarrow$ pairwise}
% \end{marginnote}

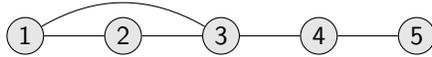
\begin{figure}
  \centering\vspace{-0.25cm}
      \begin{tikzpicture}[-, auto,
      main node/.style={circle,inner
        sep=2pt,fill=gray!20,draw,font=\sffamily}]

      \node[main node,rounded corners] (1) {1};
      \node[main node,rounded corners] (2) [right =0.8cm of 1] {2};
      \node[main node,rounded
      corners] (3) [right =0.8cm of 2] {3};
      \node[main node,rounded
      corners] (4) [right =0.8cm of 3] {4}; \node[main node,rounded
      corners] (5) [right =0.8cm of 4] {5};
            % just for vertical spacing of caption
      \node[main
      node,rounded corners,draw=none,fill=none] (6) [below =.0cm of 4]
      {};

      \path[every
      node/.style={font=\sffamily\small}]
      (1) edge node {} (2)
      (1) edge [bend left] node {} (3)
      (2) edge node {} (3)
      (3) edge node {} (4)
      (4) edge node {} (5);
    \end{tikzpicture}
  \caption{An undirected graph with five nodes.}
  \label{fig:ug:exampleMP}
\end{figure}

\begin{example}
  \label{sec2:ex:ug}
  Let $X=(X_1,\dots,X_5)$ belong to the graphical model with the undirected graph $G$ in Figure~\ref{fig:ug:exampleMP}.
  If $X$ satisfies the pairwise Markov property, the missing edges $1-4 \notin E$ and $2-4 \notin E$ imply
  $X_4\indep X_1\mid (X_2,X_3,X_5)$ and
  $X_4\indep X_2\mid (X_1,X_3,X_5)$.  The local Markov property for node $v=4$ implies
  $X_4\indep (X_1,X_2) \mid (X_3,X_5)$.  The global
  Markov property
  explicitly requires
  many other conditional independencies,  such as for example
  $X_4\indep X_2 \mid (X_1,X_3)$ or
  $(X_4,X_5)\indep (X_1,X_2) \mid X_3$.
\end{example}

The pairwise Markov property for undirected graphical models
translates each absent edge into a
`full' conditional independence.
For this reason, the smallest undirected graph $G$ with
respect to which $X$ is pairwise Markov is also known as the
conditional independence graph of $X$.  Testing all $\binom{|V|}{2}$
pairwise full conditional independencies that might arise
in~(\ref{eq:ug:pairwiseMP}) yields
a method to estimate
this graph.
 Addressing the multiple testing issues in this
approach allows for control of false edge discoveries.  We will not
discuss this in detail but refer the reader to
\cite{MR2416818}, \cite{MR3161453}, and \cite{MR3263117}.

Under the local Markov property,  each variable $X_v$
can be optimally predicted from its neighbors $X_{\nb(v)}$,
which is used in a method known as neighborhood
selection (see Section~\ref{sec:learn-undir-graph}).  The global
Markov property on the other hand is very useful for reasoning
about conditional independence.

\medskip
The famous theorem of Hammersley and
Clifford clarifies the construction of distributions that possess the
Markov properties.
Suppose $X=(X_v:v\in V)$ has a density
with respect to a product measure $\mu=\otimes_{v\in V} \mu_v$ on
$\mathbb{R}^V$.
In applications, each $\mu_v$ is usually either Lebesgue or a counting
measure, so each $X_v$ is (absolutely) continuous or discrete.  Let
$\mathcal{C}(G)$ be the set of all {\em complete subsets} (or {\em
  cliques}) of $G$, i.e., $C\in\mathcal{C}(G)$ if $\{v,w\}\in E$
for all $v,w\in C$.
Then the distribution of $X$
is said to \emph{factorize} with respect to
$G$ if it has a density of the form
\begin{equation}
  \label{eq:ug-factorization}
  f(x) \;=\; \prod_{C\in\mathcal{C}(G)} \phi_C(x_C),
  \quad x\in\mathbb{R}^{V},
\end{equation}
where each {\em potential function} $\phi_C$ has domain
$\mathbb{R}^{C}$, and $x_C$ is the subvector
$(x_v:v\in C)$.\footnote{For a finite set $A$, the space
  $\mathbb{R}^A$ comprises real vectors of length $|A|$ with entries
  indexed by $A$.}  The potential functions need not have
a probabilistic interpretation as conditional densities.

\begin{theorem}[Hammersley-Clifford]
  Suppose $X=(X_v:v\in V)$ has a positive density with
  respect to a product measure.  Then the distribution of $X$
  factorizes with respect to $G=(V,E)$ if and only if $X$ satisfies
  the pairwise Markov property with respect to $G$.
\end{theorem}

% The Construction of Multivariate Distributions from
% Markov Random Fields
% Mark S. Kaiser and Noel Cressie

% \begin{marginnote}[]
%    \entry{Factorization}{Undirected graphical models can be specified
%      via factorizing densities}
% \end{marginnote}

A graphical model may thus be defined by
specifying families of potential functions.  If the functions are
positive, then the pairwise and global Markov properties
are equivalent.
Moreover, the global Markov property lists every conditional
independence that holds in all distributions with factorizing
densities.  This is known as completeness of the global Markov
property.  We remark that factorization of nonpositive distributions
for categorical variables is treated by \cite{geiger:2006}.
\cite{MR3052406} gives a new perspective on factorization as a
consequence of log-convexity of a set of distributions.

\begin{example}
  \label{ex:ug-gaussian}
  Suppose $X=(X_1,\dots,X_5)$ is centered and multivariate normal with
  positive definite covariance matrix $\Sigma$.  Let
  $K=(\kappa_{vw})=\Sigma^{-1}$.  Then $X$ has density
  \[
    f(x) \;=\; \frac{1}{\sqrt{(2\pi)^5\det(\Sigma)}}
    \exp\left(-\frac{1}{2} \sum_{v,w=1}^5 \kappa_{vw} x_vx_w
    \right), \quad x\in\mathbb{R}^5.
  \]
  This distribution factorizes with respect to the graph
  in Figure~\ref{fig:ug:exampleMP} if and
  only if
  $\kappa_{14}=\kappa_{15}=\kappa_{24}=\kappa_{25}=\kappa_{35}=0$.
  More generally, the Gaussian model associated with an
  undirected graph $G=(V,E)$ comprises all normal
  distributions with a positive definite inverse covariance matrix
  $K=(\kappa_{vw})\in\mathbb{R}^{V\times V}$ such that $\kappa_{vw}=0$
  when $v-w\not\in E$.

 Let $S$ be the sample covariance matrix for an i.i.d.~Gaussian
  sample of size $n$.  The maximum likelihood estimator (MLE) in the
  Gaussian graphical model
  maximizes the log-likelihood function
  \begin{equation}
    \label{eq:gaussian:loglik}
    L(K) \;=\; \log\det(K) - \text{tr}\left(SK\right)
  \end{equation}
  subject to $K$ being positive definite with zeros
  over non-edges of $G$.  Strictly speaking,
  (\ref{eq:gaussian:loglik}) is obtained by maximizing over an
  unknown mean vector, dividing out $n/2$, and omitting an additive
  constant.
  If $n>|V|$, then $L$ admits a unique maximizer with probability one,
  because
  $S$ is almost surely positive definite.  If
  $|V|\ge n$, then $S$ is singular
  and $L$ can
  be unbounded.  However, if the graph $G$ is sparse,
  then the MLE of $K$ may exist uniquely with probability one
  even if
  $n$ is much smaller than
  $|V|$; see, e.g.,~\cite{Sullivant:Gross}.
\end{example}

%%%%%%%%%%%%%%%%%%%%%%%%%%%%%%%%%%%%%%%%%%%%%%%%%%%%%%%%%%

\subsection{Directed Graphical Models}
\label{sec:dir-graph-models}

Directed acyclic graphs (DAGs) are directed graphs without directed
cycles.\footnote{While graph theory speaks of `acyclic directed
  graphs' (or `acyclic digraphs'), the phrase `directed acyclic graph'
  that leads to the catchy abbreviation `DAG' has established itself in the literature on
  graphical models.}  A random vector $X=(X_v:v\in V)$ satisfies the
\emph{local Markov property} with respect to a DAG $G$ if
$$X_v \indep X_{\nd_G(v) \setminus \pa_G(v)}\, |\, X_{\pa_G(v)}$$
for every $v \in V$.
Similarly, $X$ satisfies the
\emph{global Markov property} with respect to $G$ if
$X_A\indep X_B \mid X_C$ for all triples of pairwise disjoint subsets
$A,B,C\subset V$ such that $C$ \emph{d-separates} $A$ and $B$ in $G$,
which we denote by $A \dsep_G B\, |\, C$.  The notion of d-separation in directed graphs is more subtle
than separation in undirected graphs. We refer to Definition \ref{def:dsep} below.

If $X$ satisfies the global Markov property with respect to $G$,
then $G$ is called an
\emph{independence map} of $X$.  A DAG $G$ is a \emph{perfect map}
of $X$ if
$ A \dsep_G B \,|\, C$ if and only if
$ X_A \indep X_B \,|\, X_C $
for all pairwise disjoint sets $A,B,C \subset V$.  A perfect map thus
requires the global Markov property and its reverse implication,
known as \emph{faithfulness}%
\label{def:perfect map}.
The assumption that $G$ is a perfect map is
important for the structure learning methods that we discuss in Section \ref{sec:learn-direct-graph}.

% \begin{marginnote}
%   \entry{Perfect map}{Conditional independencies in distribution exactly match d-separations in graph}
% \end{marginnote}

Suppose $X$ has a density with respect to a product measure.  Then the
distribution of $X$ \emph{factorizes} according to a DAG $G=(V,E)$ if it
has a density of the form
\begin{align}
   f(x) = \prod_{v\in V} f(x_v \,|\, x_{\pa_G(v)} ), \label{eq:DAG-factorization}
\end{align}
where the $f(x_v \,|\, x_{\pa_G(v)} )$ are conditional densities with
$f(x_v \,|\, x_\emptyset ) =f(x_v)$. The global and local Markov
properties are equivalent, and under the assumed existence of a
density they are equivalent to the factorization property \citep{Verma88}.
 The
global Markov property is also complete, i.e., it states all conditional
independencies that are implied by the factorization.
% \begin{marginnote}[]
%    \entry{Markov properties for directed graphs}{global $\Leftrightarrow$ local $\Leftrightarrow$ factorization}
% \end{marginnote}

A directed graphical model for a DAG $G$ can be specified by
conditional densities $f(x_v \,|\, x_{\pa_G(v)} )$, $v\in V$.  If these do not
share common parameters, the likelihood function of the model
factorizes into $|V|$ local likelihood functions, and the MLEs of
$f(x_v\,|\,x_{\pa_G(v)})$, $v\in V$, can be computed separately.  For
categorical data,
this amounts to computing empirical frequencies for
conditional probability tables.  For multivariate Gaussian data, the
conditional means and variances are found via linear regression
of each $X_v$ on $X_{\pa_G(v)}$.
% \begin{marginnote}[]
%    \entry{ML estimation}{Separately for each variable given its parents}
% \end{marginnote}

\medskip We now define
d-separation.  In a DAG $G=(V,E)$, nodes $v$ and $w$ are adjacent if
$v\to w\in E$ or $w\to v\in E$, and a \emph{path}
is a sequence of distinct nodes in which
successive nodes are adjacent.
If
$\pi=(v_0,v_1,\dots,v_k)$
is a path, then
 $v_0$ and $v_k$ are the \emph{endpoints} of the path.  A non-endpoint $v_i$ is
a \emph{collider} on $\pi$ if
$v_{i-1}\to v_i \leftarrow v_{i+1}$ is a subpath of $\pi$.
  Otherwise, $v_i$ is a
\emph{non-collider} on $\pi$.  If every edge on $\pi$ is of the form
$v_{i-1}\to v_i$, then $v_0$ is an \emph{ancestor} of $v_k$ and $v_k$ is a
\emph{descendent} of $v_0$.  We write $\adj_G(v)$, $\an_G(v)$ and $\de_G(v)$ for the sets of
adjacent nodes, ancestors and descendants of $v$ in $G$, respectively. We use the convention
that $v$ is an ancestor and descendant of itself, and apply the notions
disjunctively to sets, e.g.,~$\an_G(C)=\cup_{v\in C}\an_G(v)$. Finally, we define
$\nd_G(C) = V\setminus \de_G(C)$.

\begin{definition}
  \label{def:dsep} Two nodes $v$ and $w$ in a DAG $G = (V, E)$ are
  \emph{d-connected} given $C\subseteq V \setminus\{v, w\}$ if $G$
  contains a path $\pi$ with endpoints $v$ and $w$ such that (i) all
  colliders on $\pi$ are in $\an_G(C)$, and (ii) no non-collider on
  $\pi$ is in $C$.  Generalizing to sets, two disjoint subsets
  $A, B \subset V$ are d-connected given
  $C\subseteq V\setminus (A\cup B)$ if there are two nodes $v\in A$
  and $w\in B$ that are d-connected given $C$.  If this is not the
  case, then $C$ \emph{d-separates} $A$ and $B$.
\end{definition}

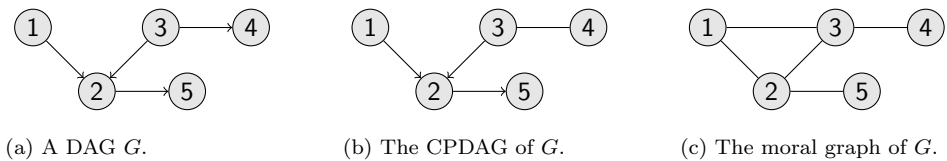
\begin{figure}
  \centering\captionsetup[subfigure]{width=1.5in}
  \subfloat[A DAG $G$.]{
    \label{fig:DAG example 1-a}
    \begin{tikzpicture}[auto,
      main node/.style={circle,inner
        sep=2pt,fill=gray!20,draw,font=\sffamily}]
      
      \node[main node,rounded corners] (1) {1};
      \node[main node,rounded corners] (2) [below right =0.7cm of 1] {2};
      \node[main node,rounded
      corners] (3) [above right =0.7cm of 2] {3};
      \node[main node,rounded
      corners] (4) [right =0.7cm of 3] {4};
      \node[main node, rounded corners] (5) [right=.7cm of 2] {5};

      \draw[->] (1) edge (2);
      \draw[->] (3) edge (2);
      \draw[->] (3) edge (4);
      \draw[->] (2) edge (5);
    \end{tikzpicture}}
    \hspace{0.85cm}
    \subfloat[The CPDAG of $G$.]{
      \label{fig:DAG example 1-c}
      \begin{tikzpicture}[auto,
      main node/.style={circle,inner
        sep=2pt,fill=gray!20,draw,font=\sffamily}]

      \node[main node,rounded corners] (1) {1};
      \node[main node,rounded corners] (2) [below right =0.7cm of 1] {2};
      \node[main node,rounded
      corners] (3) [above right =0.7cm of 2] {3};
      \node[main node,rounded
      corners] (4) [right =0.7cm of 3] {4};
      \node[main node, rounded corners] (5) [right=.7cm of 2] {5};

      \draw[->] (1) edge (2);
      \draw[->] (3) edge (2);
      \draw[-] (3) edge (4);
      \draw[->] (2) edge (5);
   \end{tikzpicture}}
      \hspace{0.85cm}
    \subfloat[The moral graph of $G$.]{
    \label{fig:DAG example 1-b}
    \begin{tikzpicture}[auto,
      main node/.style={circle,inner
        sep=2pt,fill=gray!20,draw,font=\sffamily}]

      \node[main node,rounded corners] (1) {1};
      \node[main node,rounded corners] (2) [below right =0.7cm of 1] {2};
      \node[main node,rounded
      corners] (3) [above right =0.7cm of 2] {3};
      \node[main node,rounded
      corners] (4) [right =0.7cm of 3] {4};
      \node[main node, rounded corners] (5) [right=.7cm of 2] {5};

      \draw[] (1) edge (2);
      \draw[] (2) edge (3);
      \draw[-] (3) edge (4);
      \draw[-] (1) edge (3);
      \draw[-] (2) edge (5);
    \end{tikzpicture}}
  \caption{An example to illustrate Markov properties, the CPDAG,
      and the moral graph.}
   \label{fig:DAG example 1}
\end{figure}

\begin{example}\label{ex:d-sep DAGs}
  Let
  $X=(X_1,\dots,X_5)$ belong to the graphical model with the DAG $G$
  in Figure \ref{fig:DAG example 1-a}. We see that node 2 is a collider
  on the path $1\to 2 \leftarrow 3 \to 4$, while it is  a non-collider on
  $1\to 2 \to 5$.  For node $4$, the local Markov property requires
  $X_4 \indep (X_1,X_2,X_5) \,|\, X_3$.  Since $1$ and $4$ are d-connected
  given $C=\{2\}$, $C=\{5\}$ and $C=\{2,5\}$, but d-separated given
  any other subset of $\{2,3,5\}$, the global Markov property
  requires
  $X_1 \indep X_4 \,|\, X_{C'}$ for any such other subset $C'$ of
  $\{2,3,5\}$.  We observe that, in contrast to separation in
  undirected graphs, d-separation in a DAG is not monotonic in the
  sense that $A\dsep_G B \,|\, C$ does not imply that
  $A\dsep_G B \,|\, C'$ for sets $C'\supsetneq C$.  Note also that
  there does not exist an undirected graph that encodes the same
  conditional independencies as the DAG in Figure \ref{fig:DAG example
    1-a}.  Finally, the factorization for $G$  takes the form
  $f(x) = f(x_1)f(x_2|x_1,x_3)f(x_3)f(x_4|x_3)f(x_5|x_2)$.
\end{example}

Two DAGs $G$ and $G'$ are \emph{Markov equivalent} if
$ A \dsep_G B \,|\, C$ is equivalent to $ A \dsep_{G'} B \,|\, C$.
Markov equivalent DAGs are characterized by having the same skeleton
and the same v-structures
\citep{MR1096723,VermaPearl90}. A
\emph{v-structure} is a triple of nodes $u\to v\leftarrow w$ with $u$
and $w$ not adjacent.  Each Markov equivalence class can be
represented by a \emph{completed partially directed acyclic graph}
(CPDAG) that may have directed and undirected edges
\citep[e.g.,][]{AnderssonEtAl97,MR2188675}. A CPDAG has edge $v\to w$
if and only if the edge $v\to w$ is common to all DAGs in its
equivalence class.  If the class contains a DAG with
$v\to w$ and a DAG with $v\leftarrow w$, then the CPDAG has the
undirected edge $v-w$.  The
DAG $G$ in Figure \ref{fig:DAG example 1-a} is
Markov equivalent to exactly one other DAG,
obtained by replacing the
edge $3\to 4$ by $3\leftarrow 4$.  The CPDAG
of $G$
is shown in Figure
\ref{fig:DAG example 1-c}.
% \begin{marginnote}[]
%   \entry{Markov equivalent DAGs}{Same skeleton and same v-structures}
%   \entry{CPDAG}{Represents a Markov equivalence class of DAGs}
% \end{marginnote}

\medskip The \emph{skeleton} of a (partially) directed graph is the
undirected graph obtained by replacing all edges by undirected edges.
The \emph{moral graph} $G^m$ of a DAG $G$ is constructed by first
shielding all v-structures and then taking the skeleton of the
resulting graph.  Shielding a v-structure $u\to v\leftarrow w$ means
adding an edge between nodes $u$ and $w$.  Figure \ref{fig:DAG example
  1-b} shows the moral graph of the DAG in Figure \ref{fig:DAG example
  1-a}.  It is easy to see that if $X$ satisfies the factorization
property for directed graphs with respect to $G$, then $X$ satisfies
the factorization property for undirected graphs with respect to the
moral graph $G^m$.
Hence, if $G$ is a perfect map of $X$, then $G^m$ is the conditional
independence graph of $X$. This implies that the skeleton of a DAG (or CPDAG) is a subgraph
of its corresponding conditional independence graph.

\medskip The graphical model associated with a DAG $G=(V,E)$ can also
be thought of as a structural equation model \citep{Bollen89}.
Indeed, if $\epsilon = (\epsilon_v: v\in V)$ is a vector of
independent random noise variables and $g_v$ are measurable functions,
then the random vector $X=(X_v:v\in V)$ given by
\begin{equation}\label{eq:SEM-eqns}
   X_v \;=\; g_v(X_{\pa_G(v)}, \epsilon_v), \quad v \in V,
\end{equation}
is Markov with respect to $G$.  Conversely, if $X$ is Markov with
respect to $G$, then there are independent variables $\epsilon_v$ and
functions $g_v$ such that~(\ref{eq:SEM-eqns}) holds.

\begin{example}\label{ex: SEM}
  If all functions $g_v$ are linear and the $\epsilon_v$ are normal random variables,
  then we may assume that
  (\ref{eq:SEM-eqns}) takes the form $X=BX+\epsilon$, where the
  matrix $B=(\beta_{vw})\in\mathbb{R}^{V\times V}$ has $\beta_{vw}=0$
  if $w\not\in\pa_G(v)$. The solution $X=(I-B)^{-1}\epsilon$ follows
  a multivariate normal distribution with
  covariance matrix $\cov(X) = (I-B)^{-1} \cov(\epsilon) (I-B)^{-T}$.
  Here, $I-B$ is invertible since $\det(I-B)=1$ by acyclicity of $G$.
\end{example}

Structural equation models, and thus also directed graphical models,
admit a natural causal interpretation.
To this end, one views the equations
in~(\ref{eq:SEM-eqns}) as specifying an assignment mechanism, which is
clarified by writing
\begin{align}\label{eq:SEM}
   X_v \leftarrow g_v(X_{\pa_G(v)}, \epsilon_v), \quad v \in V.
\end{align}
The variables in $X_{\pa_G(v)}$ are then treated as direct causes of
$X_v$, meaning that changes in $X_{\pa_G(v)}$ may lead to changes in
$X_v$, but not the other way around. This interpretation allows statements
about the distribution of $X$ under experimental interventions. In particular,
interventions
to the system can be modeled by changing the structural equations for
precisely those variables that are affected by the intervention
\citep{Pearl09}.
% \begin{marginnote}[]
%   \entry{Structural equation models}{Causal interpretation of DAGs
%     by treating equations as assignment mechanism
%    }
% \end{marginnote}

\begin{example}\label{ex: SEM-causal}
  The structural equation model for the DAG in Figure \ref{fig:DAG
    example 1-a} postulates that
  \begin{align*}
    X_1 &\leftarrow g_1(\epsilon_1),
    & X_3&\leftarrow g_3(\epsilon_3),
    & X_2 &\leftarrow g_2(X_1,X_3,\epsilon_2),
    & X_4&\leftarrow g_4(X_3,\epsilon_4),
    & X_5&\leftarrow g_5(X_2,\epsilon_5).
  \end{align*}
  We now consider an intervention on $X_2$, where $X_2$ is generated
  as an independent draw from the distribution of
  $\epsilon_2$. Denoting the post-intervention variables by $\tilde X  = (\tilde X_v: v\in V)$, the distribution of $\tilde X$ is induced by the equation system
  \begin{align*}
    \tilde X_1 &= g_1(\epsilon_1),
    & \tilde X_3&= g_3(\epsilon_3),
    & \tilde X_2 &= \epsilon_2,
    & \tilde X_4&= g_4(\tilde X_3,\epsilon_4),
    & \tilde X_5&= g_5(\tilde X_2,\epsilon_5).
  \end{align*}
  The post-intervention DAG $\tilde G$ is obtained from $G$ by removing the edges $1\to 2$ and $3\to 2$.
\end{example}

Thus, the causal
interpretation of a DAG allows predictions in changed environments,
and hence the estimation of causal effects.  These ideas can be
combined with the structure learning methods that we discuss in
Section \ref{sec:learn-direct-graph} to estimate (bounds on) causal
effects from observational data
\citep{MaathuisKalischBuehlmann09, MaathuisEtAl10-NatureMethods,
  PerkovicEtAl15, NandyMaathuisRichardson16}.

%%%%%%%%%%%%%%%%%%%%%%%%%%%%%%%%%%%%

\section{Learning Undirected Graphical Models}
\label{sec:learn-undir-graph}

Our treatment of learning undirected graphical models begins with the special
case of trees.  We then move on to generally applicable greedy search
and $\ell_1$ penalization methods as well as techniques that avoid
traditional assumptions of Gaussianity for continuous observations.

\subsection{Chow-Liu Trees and Forests}
\label{sec:chow-liu-trees}

A tree $G=(V,E)$ is an undirected graph with a unique path between any two nodes, and thus
$|E| = |V|-1$.
\cite{ChowLiu:1968} showed that one can
efficiently find a tree-structured distribution that optimally
approximates a given distribution.  In the present context, we may
view their algorithm as outputting a maximum likelihood (ML) estimate
of a conditional independence tree.

Due to convenient factorizations, computation in tree-based graphical
models is particularly tractable
\citep{MAL-001}.
Indeed, if the distribution of a random vector $X=(X_v:v\in V)$
factorizes with respect to a tree, then the joint density factorizes
as
\begin{equation}
  \label{eq:tree:factor}
  f(x) \;=\; \prod_{v-w\in E}
    \frac{f_{vw}(x_v,x_w)}{f_v(x_v)f_w(x_w)}\prod_{v\in V}
    f_v(x_v).
\end{equation}
Here, $f_v$ and $f_{vw}$ are the marginal densities of $X_v$ and
$(X_v,X_w)$, respectively.  Formula~(\ref{eq:tree:factor}) is a special case of a
more general result exemplified in~(\ref{eq:ug:decomposition}).
It coincides with~(\ref{eq:DAG-factorization}) when we create a
directed tree by letting edges point away from one arbitrarily
selected node.

Suppose for a moment that all variables are categorical, with
$X_v$ taking values in a finite set $\mathcal{X}_v$.  For a joint
state $x\in\mathcal{X}:=\prod_{v\in V}\mathcal{X}_v$, let $N(x)$ be
the number of times $x$ appears in an i.i.d.~sample of size $n$ from
the distribution of $X$.  Writing $\hat f_G(x)$ for the ML estimate of
the joint probability $f(x)$ in the graphical model given by tree
$G$, we aim to find the tree $G$ with largest maximum
log-likelihood
$
  \hat L(G) = \sum_{x\in\mathcal{X}} N(x)\log \hat f_G(x).
$

Let $\hat f_{vw}(x_v,x_w)$
and
$\hat f_v(x_v)$
be the relative frequencies of seeing the pair $(X_v,X_w)$ in state
$(x_v,x_w)$ and variable $X_v$ in state $x_v$, respectively.
The MLE $\hat f_G(x)$ is
obtained by plugging the $\hat f_{vw}$ and $\hat f_v$
into~(\ref{eq:tree:factor}).
It follows that
\[
  \frac{1}{n} \hat L(G) \;=\; \sum_{v-w\in E} I\big(\hat f_{vw}\big)
  + \text{const.},
\]
where $I(\hat f_{vw})$ is the empirical mutual information of $X_v$
and $X_w$, so
\[
  I\big(\hat f_{vw}\big) \;=\; \sum_{x_v\in\mathcal{X}_v}\sum_{x_w\in\mathcal{X}_w}
  \hat f_{vw}(x_v,x_w) \log\frac{\hat f_{vw}(x_v,x_w)}{\hat
    f_v(x_v)\hat f_w(x_w)}.
\]
Since mutual information is non-negative, the ML tree is a maximum
spanning tree for the complete graph with edge weights
$I(\hat f_{vw})$.  The maximum spanning tree can be computed efficiently using, e.g.,
Kruskal's algorithm, which adds edges $\{v,w\}$ in the order of
decreasing mutual information $I(\hat f_{vw})$ but skips edges that
create a cycle.
The ML tree is found after addition of $|V|-1$ edges.
If Kruskal's algorithm is stopped early, adding only $k$ edges, then
the output is a forest with maximum likelihood among all forests with
$k$ edges.  A forest is an undirected graph that is a union of
disconnected trees.
% \begin{marginnote}[]
%   \entry{Chow-Liu trees}{Maximum likelihood trees are
%     maximum spanning trees for a complete graph weighted by mutual
%     information}
% \end{marginnote}

The Chow-Liu method is not limited to categorical data.
Indeed,
we may formulate statistical models for the bivariate marginal
distributions $f_{vw}$, compute their ML estimates $\hat f_{vw}$ and
find a maximum weight spanning tree from their mutual informations
$I(\hat f_{vw})$.  In particular, when the marginals are taken to be
bivariate normal then the joint density $f$ is multivariate normal and
$ I(\hat f_{vw}) = -\frac{1}{2}\log(1-r_{vw}^2)$, where $r_{vw}$ is
the empirical correlation between $X_v$ and
$X_w$.
In this case, the absolute correlations
$|r_{vw}|$ can also be used as weights.

While it is an older idea, there has been renewed interest in
Chow and Liu's approach.
\cite{MR2789417,MR2813149} study which trees/forests are most
difficult to recover.
\cite{MR2786914} use
kernel density estimates in a nonparametric approach.
\cite{Edwards2010} discuss mixed categorical and
continuous data
and incorporate information criteria
into the algorithm.  The output is then a forest
because the penalties for model complexity
may yield negative edge weights.
Treating models with latent variables, \cite{friedman2002structural}
suggested a structural EM algorithm whose M-step optimizes over both
parameters and tree structure.  The Chow-Liu algorithm was also used
in methods for learning latent locally tree-like graphs by
\cite{MR3099108}.

\subsection{Greedy Search}
\label{sec:greedy-search-UG}

Beyond the realm of trees, finding a graph that maximizes a (penalized) likelihood
or information criterion is hard in a
complexity-theoretic sense \citep{Srebro01learningmarkov}.
Nevertheless, good estimates can be obtained from heuristic
techniques, such as greedy or stepwise forward/backward search.
Good implementations, e.g., as discussed by \cite{MR2905395}, evaluate
the benefit of an
edge addition or removal via local computations based on
clique-sum decompositions \citep[Chap.~3]{Lauritzen96}.

\begin{figure} %[h]
  \centering\captionsetup[subfigure]{width=2.5in}
  \subfloat[An undirected graph.]{
    \label{fig:graph-to-decompose}
    \begin{tikzpicture}[auto, main node/.style={circle,inner
        sep=2pt,fill=gray!20,rounded corners,draw,font=\sffamily}]

      \node[main node] (1) {1};
      \node[main node] (2) [below =0.8cm of 1] {2};
      \node[main node] (3) [right =0.8cm of 1] {3};
      \node[main node] (4) [below =0.8cm of 3] {4};
      \node[main node] (5) [above right =0.3cm and 0.8cm of 4] {5};
      \node[main node] (7) [below right =0.3cm and 0.8cm of 5] {7};
      \node[main node] (6) [above =0.8cm of 7] {6};
      % just for vertical spacing of caption
      \node[main
      node,rounded corners,draw=none,fill=none] (8) [below =.0cm of 4] {};

      \path[every
      node/.style={font=\sffamily\small}]

      (1) edge node {} (2)
      (1) edge node {} (3)
      (2) edge node {} (4)
      (3) edge node {} (4)
      (4) edge node {} (5)
      (3) edge node {} (5)
      (6) edge node {} (5)
      (7) edge node {} (5)
      (6) edge node {} (7)
;
    \end{tikzpicture}
    }
    \hspace{1.5cm}
    %\\[.25cm]
    \subfloat[Clique-sum of its prime components.]{
    \label{fig:prime-comps}
    \begin{tikzpicture}[auto, main node/.style={circle,inner
        sep=2pt,fill=gray!20,rounded corners,draw,font=\sffamily}]

      \node[main node] (1) {1};
      \node[main node] (2) [below =0.8cm of 1] {2};
      \node[main node] (3) [right =0.8cm of 1] {3};
      \node[main node] (4) [below =0.8cm of 3] {4};
      \node[main node] (4a) [right =0.8cm of 4] {4};
      \node[main node] (3a) [right =0.8cm of 3] {3};
      \node[main node] (4) [below =0.8cm of 3] {4};
      \node[main node] (5) [above right =0.3cm and 0.8cm of 4a] {5};
      \node[main node] (5a) [right =0.8cm of 5] {5};
      \node[main node] (7) [below right =0.3cm and 0.8cm of 5a] {7};
      \node[main node] (6) [above =0.8cm of 7] {6};
      % just for vertical spacing of caption
      \node[main node,draw=none,fill=none] (8) [below =.0cm of 4] {};

      \node (plus1)  [above right=0.25cm and 0.25cm of 4] {+};
      \node (plus2)  [right= 0.15cm of 5] {+};

      \path[every
      node/.style={font=\sffamily\small}]

      (1) edge node {} (2)
      (1) edge node {} (3)
      (2) edge node {} (4)
      (3) edge node {} (4)
      (3a) edge node {} (4a)
      (4a) edge node {} (5)
      (3a) edge node {} (5)
      (6) edge node {} (5a)
      (7) edge node {} (5a)
      (6) edge node {} (7)
;
    \end{tikzpicture}
  }
  \caption{A clique-sum decomposition.}
  \label{fig:ug:exampleDecompose}
\end{figure}
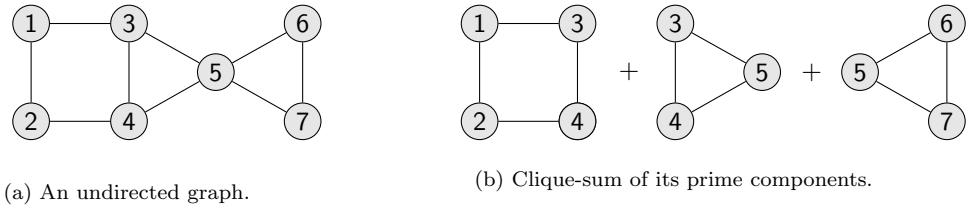

\begin{example}
  \label{ex:clique-sum}
  The graph $G$ in Figure~\ref{fig:graph-to-decompose} is a clique-sum of
  three smaller graphs, the so-called prime components shown in
  Figure~\ref{fig:prime-comps}.  If a distribution factorizes with
  respect to $G$, then its density $f$ satisfies
  \begin{equation}
    \label{eq:ug:decomposition}
    f(x)\;=\;
    \frac{f_{1234}(x_1,x_2,x_3,x_4)f_{345}(x_3,x_4,x_5)
      f_{567}(x_5,x_6,x_7)}{f_{34}(x_3,x_4)f_5(x_5)}.
  \end{equation}
  The marginal densities in the numerator correspond to the
  prime components, and those in the denominator correspond to
  the separating sets, i.e., the cliques along which the prime components
  are summed.  Suppose now that we wish to compute the likelihood
  ratio statistic comparing the given graph to the graph with edge
  $3-5$ removed, in a setting of categorical data under multinomial
  sampling.  The edge belongs to the clique $\{3,4,5\}$, Proposition 4.32 in \cite{Lauritzen96}
  implies that the
  likelihood ratio statistic can be obtained by computing with the
  data on $(X_3,X_4,X_5)$ alone.  Specifically, one computes the
  likelihood ratio of the graph $3-4-5$ with respect to the full
  triangle graph on $\{3,4,5\}$.
Analogous results exist for the
  Gaussian case and, with additional subtleties, for mixed
  discrete and
  conditionally Gaussian observations \citep[Chap.~5 and 6]{Lauritzen96}.
% \begin{marginnote}
%   \entry{Local computation}{Graph decomposition allows for efficient
%     computation of likelihood after edge addition, removal
%     and reversal}
% \end{marginnote}
\end{example}

Computation of likelihood ratios is particularly convenient for
decomposable graphs, i.e., graphs in which all prime components are
complete.  For example, trees are decomposable because their prime
components are the individual edges, and the graph in Figure~\ref{fig:graph-to-decompose}
  is non-decomposable.
For
a decomposable graph, Gaussian models as well as models for
categorical variables admit closed form MLEs \cite[Sections~4.4,
5.3]{Lauritzen96}.
These are obtained by estimating marginal
distributions (via sample means and covariances for Gaussian data,
or via empirical frequencies for categorical data) and substituting
these estimates in a factorization such
as~(\ref{eq:ug:decomposition}).
In contrast, computing MLEs for non-decomposable graphs involves
solving higher degree polynomial equation systems
\cite[Chap.~2.1]{MR2723140}.

More recently, greedy search has been applied in a framework of
neighborhood selection, in which a graph $G$ is selected by determining
the neighborhood $\nb_G(v)$ of each node $v$.  As discussed in
Section~\ref{sec:neighborhood-selection}, finding $\nb_G(v)$ often
corresponds to variable selection in a regression problem.  This
avoids the need for iterative computation of MLEs when dealing with
non-decomposable graphs, and a connection can be made to results on
forward selection methods for variable selection in high-dimensional
regression.  \cite{Jalali:2011} and \cite{MR3352509} leverage this
connection and provide theoretical guarantees for greedy search in
high-dimensional problems.  The related method of
\cite{Bresler:2015:ELI:2746539.2746631} greedily selects supersets of
the neighborhoods that are subsequently pruned.  These methods are
competitive to the $\ell_1$-regularization techniques that we discuss
next.

% \begin{marginnote}[]
%   \entry{Greedy neighborhood selection}{Neighbors found using
%     greedy search for variable selection in regression}
% \end{marginnote}

\subsection{Gaussian Models and $\ell_1$-Penalization}
\label{sec:glasso}

Gaussian models provide the starting point for most graphical modeling
of continuous observations.  As noted in Example~\ref{ex:ug-gaussian},
a Gaussian conditional independence graph can be estimated by
determining the zero entries of the inverse covariance matrix
$K\in\mathbb{R}^{V\times V}$.

Let $S=(s_{vw})$ be the sample covariance matrix for a sample of $n$
observations, with $s_{vv}>0$ for all $v\in V$ to avoid trivialities.
\cite{MR2367824} and \cite{MR2417243} proposed the graphical lasso
(\emph{glasso}) estimator
\begin{equation}
  \hat K^{\rm gl} \;=\; \arg\min_{K}\;\Big\{
  -\log\det(K) + \text{tr}\left(SK\right) + \lambda
  \|K\|_1\Big\},\label{eq:glasso}
\end{equation}
where the minimization is over positive definite matrices
$K=(\kappa_{vw})\in\mathbb{R}^{V\times V}$ and $\lambda\ge0$ is a
tuning parameter.  The objective
adds to the log-likelihood function from~(\ref{eq:gaussian:loglik}) a
multiple of the (vector) $\ell_1$-norm, i.e.,
$\|K\|_1=\sum_{v,w\in V} |\kappa_{vw}|$.  Some authors omit the
positive diagonal entries $\kappa_{vv}$ when forming the norm.  In
either case, such a regularization term induces sparsity in
$\hat K^{\rm gl}$, just as it does in lasso
regression.  The conditional independence graph is then estimated by
the graph $\hat G^{\rm gl}$ that has edge $v-w$ if and only if
$\hat \kappa^{\rm gl}_{vw}\not=0$.  For $\lambda>0$, the
minimum in~(\ref{eq:glasso}) is achieved uniquely because the
objective is strictly concave and coercive irrespective of whether $S$
has full rank. This is important in high-dimensional settings.

% \begin{marginnote}[]
%   \entry{Glasso}{Optimize $\ell_1$-penalized joint Gaussian
%     log-likelihood}
% \end{marginnote}

The coordinate-descent algorithm of \cite{glasso} is a popular method
for computation of $\hat K^{\rm gl}$; see \cite{MR3020259} for a
discussion of its properties.
Recent implementations exploit that simply thresholding the sample covariance matrix $S$ yields the connected components of $\hat G^{\rm gl}$
\citep{MR2878953,MR2913718}.
Alternative approaches for computation of $\hat K^{\rm gl}$ are
discussed in \cite{NIPS2013_4923}.
The estimation error of
$\hat K^{\rm gl}$ and the consistency of $\hat G^{\rm gl}$ in
high-dimensional problems are studied by \cite{MR2836766}.

There are several related methods to estimate a sparse covariance matrix. For example, \cite{MR2847973} minimize
$\|K\|_1$ subject to a constraint on $\|SK-I\|_\infty$.
Here, $I$ is the identity and $\|A\|_\infty$ is the maximum absolute
entry of $A$.
Minimax optimality properties can be proven for an adaptive version of
the estimator \citep{ACLIME}.

\smallskip
A conditional independence graph is sometimes expected to have
particular structure, such as
`hub' nodes with
many neighbors.
This motivated \cite{NIPS2012_4538} to consider regularization with a
sorted $\ell_1$ norm.
For tuning parameters $\lambda_1\ge\dots\ge\lambda_{p}\ge 0$, the
sorted $\ell_1$ norm of $\beta\in\mathbb{R}^p$ is
$\sum_{j=1}^{p} \lambda_j |\beta_{(j)}|$, where
$\beta_{(1)},\dots,\beta_{(p)}$ are the entries of $\beta$ listed in
descending order of absolute values, so
$|\beta_{(1)}|\ge |\beta_{(2)}|\ge\dots \ge |\beta_{(p)}|$.
\cite{NIPS2012_4538} estimate the inverse covariance
$K=(\kappa_{vw})$ by the optimal solution of
\begin{equation}
  \min_{K}\;\Big\{-
  \log\det(K) + \text{tr}\left(SK\right) + \sum_{v\in
    V}\sum_{j=1}^{|V|}\lambda_j |\kappa_{v,(j)}|\Big\}.\label{eq:defazio}
\end{equation}
Intuitively, the sorted $\ell_1$ norm allows one to more easily detect
a signal $\kappa_{vw}$ if it concerns a variable
$X_v$ with other stronger signals $\kappa_{vu}$.
Of course, the
work just described is not the only one addressing hubs
\citep[see e.g.,][]{MR3277170}.

\subsection{ Neighborhood Selection}
\label{sec:neighborhood-selection}

We may
estimate a conditional independence graph $G=(V,E)$ by estimating all
of its neighborhoods $\nb_G(v)$.  According to the local Markov
property, $X_v$ depends on the other variables only through its
neighbors $X_w$, $w\in\nb_G(v)$.  Hence, we may proceed by estimating
the conditional distribution of $X_v$ given $X_{V\setminus\{v\}}$, and
determine $\nb_G(v)$ as the index set of the variables $X_w$ on which
the estimated conditional distribution
depends.\footnote{Alternatively, we could motivate the approach by
  referring to the pairwise Markov property, from which we conclude
  that $\{v,w\}\notin E$ if the conditional distribution does not
  depend on $X_w$.}  The ideas behind this {\em neighborhood
  selection} have a longer tradition \citep{Besag1975}.  However, its
wide-spread use in graphical modeling emerged more recently when
\cite{MR2278363} tackled high-dimensional problems through a
connection to lasso regression.

\begin{example}
  \label{ex:nb-select-gaussian}
  Let $X=(X_v:v\in V)$ be multivariate normal with inverse covariance
  matrix $K=(\kappa_{vw})$.  Then the conditional distribution of
  $X_v$ given all remaining variables is normal with variance
  $1/\kappa_{vv}$ and expectation
  \[
    \E\left[X_v\,|\,X_w,\,w\not=v\right] \;=\; \sum_{w\in
      V\setminus\{v\}} \left(-\frac{\kappa_{vw}}{\kappa_{vv}} \right)
    X_w
    \;=\;\sum_{w\in\nb_G(v)} \left(-\frac{\kappa_{vw}}{\kappa_{vv}} \right)
    X_w.
  \]
  We may estimate $\nb_G(v)$ as the set of active
  covariates in a linear regression of $X_v$ on all
  other variables $X_w$, $w\not=v$. Any technique for variable
  selection could be applied.
\end{example}

\begin{example}
  \label{ex:nb-select-ising}
  In a symmetric Ising model with (real-valued) interaction parameters
  $\theta_{vw}$, all random variables $X_v$ take values in $\{-1,1\}$
  and joint probabilities have the form
  \begin{equation}
    \label{eq:ising}
    \Pr(X_v=x_v,\, v\in V) \;\propto\;
    \exp\Bigg\{\sum_{\{v,w\}\in E} \theta_{vw}x_vx_w \Bigg\}.
  \end{equation}
  By the Hammersley-Clifford theorem, the conditional independence
  graph $G=(V,E)$ of such a distribution has edge $v-w\in E$ if and
  only if $\theta_{vw}\not=0$.  Since
  \[
    \log\left( \frac{\Pr(X_v=1\,|\,X_w=x_w,\,
        w\not=v)}{1-\Pr(X_v=1\,|\,X_w=x_w,\, w\not=v)}\right) \;=\;
    \sum_{w\in\nb_G(v)} \left(2\theta_{vw}\right) x_w,
  \]
  the neighborhood $\nb_G(v)$ can be estimated as the set of active
  covariates in a logistic regression of $X_v$ on all other variables
  $X_w$, $w\not=v$.  We note that the normalizing constant
  in~(\ref{eq:ising}) is a sum over $2^{|V|}$ joint states and thus
  intractable unless $|V|$ is small.
\end{example}

Estimating each neighborhood in isolation may lead to inconsistencies
that are commonly resolved post-hoc.  Let $\widehat\nb(v)$, $v\in V$,
be the estimated neighborhoods.  If $w\in\widehat\nb(v)$ but
$v\not\in\widehat\nb(w)$, then the so-called `and'-rule excludes the edge
$v-w$ from the estimated conditional independence graph, while the `or'-rule includes such edges.

The use of $\ell_1$-regularization for neighborhood selection in
Gaussian and Ising models (Examples~\ref{ex:nb-select-gaussian} and
\ref{ex:nb-select-ising}) is studied by \cite{MR2278363} and
\cite{MR2662343}, respectively.  \cite{JMLR:v16:yang15a} treat other
exponential family models, and \cite{MR3335095} propose refinements of
the `and'/`or'-rules that
take into account the distributional type of the
nodes.
\cite{MR3180659} apply techniques for sparse additive models, in which
a conditional expectation of the form
$\E\left[X_v\,|\,X_w,\,w\not=v\right] \;=\; \sum_{w\not=v}
f_{vw}(X_w)$ is estimated using basis expansions and a group lasso
penalty that allows for zero functions as estimates of some of the
univariate functions $f_{vw}$.

% \begin{marginnote}[]
%   \entry{Neighborhood selection}{Neighbors found using
%     $\ell_1$-penalization techniques for variable selection}
% \end{marginnote}

\medskip

Neighborhood selection is related to the concept of
pseudo-likelihood, which is based on the
full conditional densities of a joint density $f$. The
log-pseudo-likelihood function is
\begin{equation}
  \label{eq:pseudo}
  L_{\rm pseudo}(f) \;=\; \sum_{v\in V} \log
  f_{v|V\setminus\{v\}}\left(x_v\,|\,x_w,\,w\not=v\right).
\end{equation}
Non-Gaussian distributions specified via the Hammersley-Clifford
theorem typically have intractable normalizing constants; recall
Example~\ref{ex:nb-select-ising}.  In contrast, the full conditionals
in the pseudo-likelihood can often be normalized.  Indeed, the
conditional probabilities for discrete $X_v$ can be normalized by
summing over the state space of $X_v$ alone as opposed to over all
joint states.  Similarly, it may be feasible to find the normalizing
constant of the full conditional density for a continuous random
variable by univariate integration.

While neighborhood selection treats the different conditional
densities $f_{v|V\setminus\{v\}}$ as unrelated, the conditionals share
parameters.  For instance, the interaction parameter $\theta_{vw}$ in
the Ising model from~(\ref{eq:ising}) appears in the conditionals for
$X_v$ and for $X_w$.  As an alternative that avoids inconsistencies
among estimated neighborhoods, we may maximize an $\ell_1$ penalized
version of $L_{\rm pseudo}$ from~(\ref{eq:pseudo}) with respect to a
symmetric interaction matrix. \cite{MR2505138} explore this
pseudo-likelihood method for Ising models
but find rather little difference with neighborhood selection.
\cite{MR3382598} give an overview of Gaussian pseudo-likelihood
methods that retain the symmetry of the inverse covariance
matrix
and address issues in the specification of a convex optimization
objective.

\medskip

A full likelihood may be expected to yield more efficient
estimators than neighborhood selection or pseudo-likelihood.  However,
under $\ell_1$ regularization, the situation is subtle as
different irrepresentability conditions are needed to ensure
consistency.
Indeed,
there are Gaussian examples in which neighborhood selection is
consistent whenever the glasso
is consistent, but in which the converse is false
\citep{MR2398362,MR2836766}.

\subsection{Score Matching}
\label{sec:score-matching}

The Hammersley-Clifford theorem allows the specification of graphical models
as interaction models in the form of an exponential family.  However, the
normalizing constants in such models are tractable only in special
cases.  The score matching approach of
\cite{MR2249836,MR2338984} is well
suited to address this challenge.  We describe the basic version
that applies to continuous observations supported on all of
$\mathbb{R}^{V}$.

% \begin{marginnote}[]
%   \entry{Score matching}{Convex quadratic loss for continuous
%     non-Gaussian graphical models}
% \end{marginnote}

Let $X=(X_v:v\in V)$ be absolutely continuous with differentiable
density $f_0$ and support $\mathbb{R}^{V}$.  Let $f$ be another density
that is twice differentiable and has support $\mathbb{R}^{V}$.  Writing
$\nabla_x$ for the gradient with respect to $x$, define the Fisher
information distance
\begin{equation}\label{divergence}
  J(f) \;=\; \int_{\mathbb{R}^{V}} f_0(x)\;\left\| \nabla_x \log f(x) -\nabla_x \log
    f_0(x) \right\|_2^2 \:dx.
\end{equation}
While it is natural to minimize an estimate of $J(\cdot)$, this approach
is complicated by the way the unknown true density $f_0$ appears
in~(\ref{divergence}).  \cite{MR2249836} circumvents this problem
using integration by parts (Stein's identity), which yields under mild
conditions that
\begin{equation}\label{scorematch}
  J(f) \;=\; \int_{\mathbb{R}^{V}} f_0(x)
  \left[\Delta_x\log
    f(x) + \frac{1}{2}\|\nabla_x \log f(x)\|_2^2\right]\:
  dx \;
  + \; \mbox{const.},
\end{equation}
where
$\Delta_x f(x) = \sum_{v} \partial^2f(x)/\partial x_v^2$
is the Laplace operator.  Writing
$S(x,f)=\Delta_x\log f(x) + \frac{1}{2}\|\nabla_x \log f(x)\|_2^2$,
a score matching estimator minimizes the
empirical loss
\[
  \hat{J}(f) = \frac{1}{n} \sum_{i=1}^n S(x^{(i)},f)
\]
for $f$ ranging over a model of interest.  Importantly, if $f$ is
only known up to a normalizing constant, then this constant cancels in
the logarithmic derivatives in $S(x,f)$.  Moreover, in an exponential
family with log-densities
$\log f(x | \theta) = \theta^T t(x) - \psi(\theta) + b(x)$ for
sufficient statistics $t(x)$, the loss $\hat{J}$ is a {convex
  quadratic} function of the natural parameter $\theta$.

\begin{example}
  Consider the family of centered multivariate normal
  distributions, parameterized by their inverse covariance matrices $K$.
  Then,
  \begin{equation}\label{eq:gauss-score-match}
    \hat{J}(K) \;=\;\frac{1}{2} \text{trace}(K^2S)-\text{trace}(K)
    \;=\; \sum_{v\in V}\frac{1}{2}\kappa_v^TS\kappa_v - \kappa_v^T e_v,
  \end{equation}
  where $S$ is the sample covariance matrix of $X$, $\kappa_v$ is the $v$-th column of $K$ and $e_v$ is the $v$-th
  canonical basis vector of $\mathbb{R}^{V}$.  If $S$ is invertible then
  the score matching estimator equals the MLE $\hat K = S^{-1}$.
  However, for submodels that constrain $K$ to lie in a linear
  subspace, the two estimators generally differ. The score matching
  estimator need not be asymptotically efficient, as \cite{MR3338335}
  show in the context of symmetry constraints in graphical models.
\end{example}

Closed form score matching
estimators are available for any pairwise interaction model
\begin{equation}
  \label{expfam:pairwise:general}
  \log f(x | \theta) \;=\; \sum_{a=1}^A\sum_{v\not=w}
  \theta^{(a)}_{vw} \,t^{(a)}_{vw}(x_v,x_w)
  \:+\:\sum_{l=1}^L\sum_{v}
  \theta^{(l)}_v t^{(l)}_v(x_v) \,-\, \psi(\theta) \,+\, b(x), \quad x\in
  \mathbb{R}^{V},
\end{equation}
for which $X_v\indep X_w\mid X_{V\setminus\{v,w\}}$ if and only if
$\theta^{(a)}_{vw}=0$ for all $a=1,\dots,A$.  Sparse estimates of the
interaction matrices $(\theta^{(a)}_{vw})$ can be obtained by adding
an $\ell_1$ or group lasso penalty to the loss $\hat J$.  The
resulting estimators of conditional independence graphs
are studied
by \cite{lin2015high} who also treat nonnegative observations, by
\cite{janofsky} who proposes a nonparametric exponential series
approach, and by \cite{NIPS2015_6006} who consider
infinite-dimensional exponential families.  For Gaussian models,
$\ell_1$-regularized score matching is a simple but state-of-the-art
method.  It coincides with the method of \cite{MR3306432}.

\subsection{Semiparametric and Robust Methods}
\label{sec:semiparametric}

Traditionally, methods for continuous observations rely heavily on
Gaussian models.  A problem of obvious interest is to provide methods
for non-Gaussian observations.  We already mentioned a number of
such methods
and comment here on two other lines of work.

From a perspective of robustness, several authors explored the use of
elliptical distributions \citep{MR2860334,MR3286922,MR3208335}. \cite{MR2840186} consider the
special case of $t$-distribution models in which a
Gaussian
random vector is observed under scaling with a single random divisor.
They also propose non-elliptical
`alternative' $t$-distributions, resulting from dividing the
different components of a latent
Gaussian vector by
independent scalars.  Different divisors are useful
for high-dimensional data with outliers in many observations, but
where each observation only has a small number of corrupted
entries.

% \begin{marginnote}[]
%   \entry{Elliptical distributions}{Robust inference}
%   \entry{Gaussian copula models}{Efficient computation based on rank
%     correlations}
% \end{marginnote}

In a different vein, \cite{MR2563983} propose semiparametric methods
based on Gaussian copula models.  Here, the observation
$X=(X_v:v\in V)$ satisfies $X_v=h_v(Z_v)$ for a
Gaussian
random vector $Z=(Z_v:v\in V)$ and strictly
increasing
functions $h_v:\mathbb{R} \to \mathbb{R}$.  Since the $h_v$ are
deterministic and one-to-one, $X$ has the same Markov properties as
$Z$.  \cite{MR3059084} and \cite{MR3097612} observe that
efficient estimation in the copula models can be based on Kendall's
$\tau$ and Spearman's $\rho$.  Indeed, for strictly increasing $h_v$,
the observation $X$ and the latent vector $Z$ have the same rank
correlations.  One may thus apply Gaussian methods after estimating
the latent Gaussian correlation matrix by suitably transformed
pairwise estimates of $\tau$ or $\rho$.  In the data analysis in
Example~\ref{ex:Marbach}, we applied this idea in conjunction with
Gaussian neighborhood selection from
Section~\ref{sec:neighborhood-selection}.

It is noteworthy that Kendall's $\tau$ also provides a simple way to
fit copula models based on elliptical distributions
\citep{liu2012transelliptical}.  Extensions to mixed discrete and
continuous data are treated by \cite{RSSB:RSSB12168}.  Avoiding the
assumption of a Gaussian copula, \cite{2014arXiv1412.8697Y} use
coarser data summaries than ranks to handle settings in which the full
conditionals are generalized linear models with unknown base measure.

\subsection{Tuning Parameter Selection}
\label{sec:tuning-parameter}

Many of the aforementioned
methods
depend on a tuning parameter.  Varying this parameter typically yields
a useful ranking of edges.  However, it may also be desirable to
select a single tuning parameter, for example by
optimizing
information criteria such as the Bayesian information criterion (BIC).
However,
in problems with a large number of variables $|V|$,
the BIC tends to yield
overly dense graphs.
\cite{Foygel:2010}, \cite{Gao2012} and \cite{MR3326135} address this
issue by adapting ideas from sparse high-dimensional regression.
Via a multiplicity-correcting prior, the BIC penalty is made dependent
on $\log |V|$.

Another useful approach is stability selection
\citep{MR2758523,Samworth2013}.   This method records how often
each edge is selected across random subsamples and different tuning parameters,
and selects
those edges for which
there exists a tuning parameter so that
the subsample selection frequency exceeds a specified threshold.  The
method aims at avoiding false positives.
A resampling technique that
seeks to avoid false negatives was proposed by \cite{Liu:2010}.  This
method chooses the least amount of penalization for which graph
estimates are suitably stable across subsamples.

Finally, we note that there are methods that aim to reduce or
eliminate dependence on tuning parameters; see
\cite{2014arXiv1410.7279L} and references therein.

%%%%%%%%%%%%%%%%%%%%%%%%%%%%%%%%%%%%

\section{Learning Directed Graphical Models}
\label{sec:learn-direct-graph}

We now consider learning the structure of a
directed graphical model.  Textbooks on this problem
include \cite{SpirtesEtAl00}, \cite{Neapolitan04}, and
\cite{KollerFriedman09}.
Throughout, we assume that the DAG $G=(V,E)$ is a
perfect map of $X=(X_v:v\in V)$, and that we observe $n$ i.i.d.\ copies of $X$. The observed data are denoted by $\mathbf{x}$.

In general, $G$ is not identifiable from the distribution of
$X$, but we can identify its Markov equivalence class, or
equivalently, its CPDAG. Thus, many structure learning methods aim to
learn the CPDAG. We treat exact score-based search in problems of
moderate dimensionality
and
review more broadly applicable methods based on greedy search
or conditional independence tests,
as well as hybrids
of these two approaches.
Finally, we discuss methods that
impose additional assumptions that allow identification of the
DAG.

\subsection{Exact Score-Based Search}
\label{sec:learn-direct-exact}

Score-based approaches learn a DAG by determining the graph $G$ that
optimizes a specified score $Q(G,\mathbf{x})$.  Typically $Q$ is a penalized likelihood
score, for example the BIC. Such scores are often \emph{decomposable},
meaning that
$Q(G,\mathbf{x}) = \sum_{v\in V} q(v \,|\, \pa_G(v), \mathbf{x})$,
where the summands
are local scores of
each node $v$ given its parents.  Scores such as the BIC are also
\emph{score-equivalent},
meaning that
$Q(G,\mathbf{x})=Q(G',\mathbf{x})$ if $G$ and $G'$ are Markov
equivalent.

Finding an optimal DAG, or possibly CPDAG, is hard due to the large
search space and the acyclicity constraint.  For instance, there are
over $10^{36}$ (labeled) DAGs on 14 nodes.  Nevertheless, for
decomposable scores, an exact search is feasible more broadly than one
might expect.  Different approaches to exact search include branch and
bound methods \citep[e.g.,][]{DeCamposEtAl09}, partial order covers
\citep[e.g.,][]{ParviainenKoivisto09}, and as we discuss in more
detail below, dynamic programming and integer linear programming.

\cite{SilanderMyllymaki06} propose
an elegant dynamic programming approach.
It leverages the fact that a best DAG for a variable set
$W\subseteq V$ can be thought of as a best sink $s \in W$, with best
parents among subsets of $W\setminus \{s\}$, and a best DAG for
$W\setminus \{s\}$.  Using dynamic programming and starting from the
singleton sets, a best sink can be found for all subsets
$W\subseteq V$.  Backtracking then yields an ordering of the nodes that is
compatible with a best DAG on $V$.
Given this ordering,
one can use regression to select parents for each node from its
predecessors in the ordering, yielding a best DAG on $V$.
% \begin{marginnote}[]
%    \entry{Exact score-based search}{Dynamic programming approaches}
% \end{marginnote}
While the computational and memory requirements are exponential in
$|V|$, this approach is feasible for problems with up to roughly 30 nodes.

More recently, authors such as \cite{JaakkolaEtAl10}, \cite{Cussens13}
and \cite{MR3178416} suggested integer linear programming approaches,
in which the search over graph structures is formulated as a linear program
over a polytope $\mathcal{P}$ representing DAGs.  For instance, the vertices of
$\mathcal{P}$ may be taken to be sparse binary vectors
$\eta=(\eta_1,\dots,\eta_{|V|})$, where each $\eta_v$ is of length
$2^{|V|-1}$, which is the number of possible parent sets. If
$\pa(v)=s_v$, then we set $\eta_v(s_v)=1$ and all other entries of
$\eta_v$ are zero.  The polytope $\mathcal{P}$ is then the convex hull of all
binary vectors $\eta$ that correspond to DAGs.
A key property of $\mathcal P$ is that cyclic graphs lie outside of $\mathcal P$.
Using the notation $\eta$ also
for interior points of $\mathcal{P}$, the structure learning problem can be
cast as
% \begin{marginnote}[]
%   \entry{Exact score-based search}{Integer linear programming approaches}
% \end{marginnote}
\begin{align*}
   \max\, \sum_{v\in V} \sum_{s_v \subseteq V\setminus\{v\}} \eta_v(s_v) q(v \,| \, s_v,\mathbf{x})\quad \text{s.t.} \quad \eta \in \mathcal{P}.
\end{align*}
The complexity of this linear program is in the facets that define the
polytope $\mathcal{P}$, and practical algorithms are based on
relaxations of $\mathcal{P}$.

\subsection{Greedy Score-Based Search}
\label{sec:learn-direct-greedy}

For large graphs, exact search is infeasible, and one can turn to greedy search. A well-known algorithm of this type is the Greedy Equivalence Search (GES) algorithm of \cite{Chickering02}.
Given a starting graph (often the empty graph) and a score, GES performs a greedy search on the space of CPDAGs. The algorithm performs a forward phase in which edges are added, and a backward phase in which edges can be removed.
Efficient implementations use local computations to evaluate the benefit of an added or deleted edge, using the decomposability of the score. The forward phase tends to take much longer in practice than the backward phase.
% \begin{marginnote}[]
%   \entry{GES}{Greedy score-based search on the space of CPDAGs}
% \end{marginnote}

Due to the greedy search, GES will typically not find the global
optimum of the score given data $\mathbf{x}$ with sample of size $n$. Remarkably, however, \cite{Chickering02} showed
that
GES does
find the global optimum with probability converging to 1 as
$n\to \infty$, if the score is decomposable, score-equivalent and
consistent.
The consistency property ensures that the true CPDAG gets the highest score with probability converging to 1 as $n\to \infty$.
An important ingredient of Chickering's proof is the fact that the forward phase outputs an independence map of $X$, with probability converging to 1 as $n\to \infty$.
% \begin{marginnote}[]
%    \entry{GES}{Consistency despite greedy search}
% \end{marginnote}

Although the number of added and deleted edges in GES is polynomial in
the number of nodes $|V|$, the number of performed score evaluations
can be exponential in $|V|$. \cite{ChickeringMeek15} show that the
backward phase of GES can be made polynomial for sparse graphs.  With
a naive forward phase that simply gives the complete graph (which is
trivially an independence map), this yields a polynomial time
algorithm for sparse graphs.

The consistency result of \cite{Chickering02} is for a classical setting where $|V|$ is fixed and $n \to \infty$. \cite{VandeGeerBuehlmann13} consider a high-dimensional setting where $|V|$ is allowed to grow with $n$. They show that the global optimum of the $\ell_0$-penalized likelihood score is consistent, but they do not propose an algorithm to find the global optimum. \citet{NandyEtAl16-ARGES} show a first high-dimensional consistency result for GES.

\subsection{Constraint-Based Methods}
\label{sec:learn-direct-constraint-based}

Constraint-based methods seek to find a DAG that is compatible with
the conditional independencies seen in the given data set.
We begin the discussion of such methods by focusing on the estimation
of the skeleton of the DAG (or, equivalently, its CPDAG), which is
typically the most computationally intensive step.  %

If $G$ is a perfect map of $X$, then $v$ and $w$ are adjacent in $G$
if and only if $X_v$ and $X_w$ are conditionally dependent given
$X_C$
for all $C \subseteq V \setminus \{v,w\}$.  A naive approach
thus determines if $v$ and $w$ are adjacent by testing all $2^{|V|-2}$
conditional independencies, as done by the SGS or IC algorithms
\citep{VermaPearl90,SpirtesEtAl00}.  In contrast, the PC algorithm of
\cite{SpirtesEtAl00} limits the number of
 conditional
independence tests, by using that $v$ and $w$ are not adjacent in a
DAG $G$ if and only if $X_v \indep X_w \,|\, X_{\pa_G(v)}$ or
$X_v \indep X_w \,|\, X_{\pa_G(w)}$. Since we do not know the parent
sets (this would mean knowing the DAG), the PC algorithm ensures that
the following conditional independencies are tested:
$X_v \indep X_w \,|\, X_C$
for all $C \subseteq \adj_G(v)$ and all
$C \subseteq \adj_G(w)$, and removes the edge if a conditional
independence is found. This still appears infeasible, however, since
we do not know $\adj_G(v)$ and $\adj_G(w)$. The PC algorithm
circumvents this problem by always working with supergraphs of the
true skeleton, and testing conditional independencies given subsets of
adjacency sets in these supergraphs.
% \begin{marginnote}[]
%    \entry{Constraint-based methods}{Reverse engineer the CPDAG based
%      on conditional independencies inferred from data}
% \end{marginnote}

Concretely, the PC algorithm starts with a complete graph on $V$.  It
then tests marginal independence for all pairs of nodes, and removes
an edge if an independence is found. Next, for every pair of nodes
that are still adjacent, it tests conditional independence given
all subsets of cardinality 1 of the adjacency sets of the two nodes. The algorithm
removes an edge if a conditional independence is found. The
algorithm continues in this manner, each time increasing the
cardinality of the conditioning set by 1, until the cardinality of the conditioning sets
exceeds $\max_{v\in V} |\adj_{G'}(v)|-1$, where $G'$ is the current
state of the skeleton.  At this point, all required conditional
independencies have been considered, and the skeleton is found.
% \begin{marginnote}[]
%   \entry{PC algorithm}{Limit the number of conditional independence
%     tests and keep conditioning sets small}
% \end{marginnote}

In a second phase, the PC algorithm post-processes the results of the
conditional independence tests and learns the v-structures, as
illustrated in Example \ref{ex:PC algo} below.  Finally, the algorithm
applies a few simple orientation rules to orient some of the remaining
edges, based on the fact that one may not create directed cycles or
new v-structures. If the true conditional independencies are used as
input for the PC algorithm, then the final output is the CPDAG of the
underlying DAG $G$.  We note that for graphs with bounded degree,
i.e., a bound on $\max_{v\in V} |\adj_G(v)|$, the PC algorithm has a running
time that is polynomial in the number of variables.  The running time
depends exponentially on the degree.

\begin{example}\label{ex:PC algo}
  Suppose we have three random variables $X_u,X_v,X_w$ with
  $X_u\indep X_w$ as the sole conditional independence.  The PC
  algorithm starts with a complete undirected graph on $\{u,v,w\}$.
  When testing marginal independencies, it removes the edge $u-w$
  after finding $X_u \indep X_w$.  No other conditional independencies
  are found, so that $u-v-w$ is the final skeleton.  Next, the
  algorithm detects that $v$ is a collider on the path $(u,v,w)$,
  since otherwise $u \dsep_G w$ is violated. Hence, the DAG (and
  corresponding CPDAG) is $u\to v \leftarrow w$.

  The same skeleton is found if the only conditional
  independence is $X_u \indep X_w \,|\, X_v$.  Then, however, $v$
  must be a non-collider on the path $(u,v,w)$ to ensure that
  $u \dsep_G w \,|\, v$
  holds. This leaves three
  possible DAGs, namely, $u\to v\to w$, $u \leftarrow v \leftarrow w$
  or $u \leftarrow v \to w$, which form a Markov equivalence class
  represented by the CPDAG $u-v-w$.
\end{example}

In practice,
conditional independencies need to be tested based on data.  Standard
tests are available for multivariate Gaussian and multinomial data.
Usually, all tests are performed at the same significance level
$\alpha$, and an edge is removed if the null hypothesis of
(conditional) independence is \emph{not} rejected.  Here, $\alpha$
does not control an overall Type I error.  Rather, it is a tuning
parameter that typically gives sparser graphs for smaller values.

High-dimensional consistency of the PC algorithm for certain sparse
Gaussian DAGs is shown in
\cite{KalischBuehlmann07}.
\cite{HarrisDrton13} generalize this result to Gaussian copulas.
\cite{ColomboMaathuis14} observe that the output of the PC algorithm
can depend strongly on the variable ordering. They provide
order-independent versions of the algorithm that are again consistent
in high-dimensional settings.
Consistency of the PC algorithm rests on the assumption that the
underlying DAG is a perfect map.  This assumption is studied in depth by
\cite{UhlerEtAl13}, who show that it can be restrictive.

\subsection{Hybrid Algorithms}
\label{sec:learn-direct-hybrid}

Hybrid algorithms combine ideas from constraint-based and
score-based methods, by employing a greedy search over a restricted
space, often determined using conditional independence tests.
An example is the Max-Min Hill Climbing (MMHC) algorithm
\citep{TsamardinosEtAl06}.  Hybrid algorithms
scale well with respect to the number of variables and exhibit
good estimation performance \citep{TsamardinosEtAl06,
  NandyEtAl16-ARGES}.

The theoretical properties of hybrid algorithms are less well studied
than those of purely score- or constraint-based algorithms.
\cite{NandyEtAl16-ARGES} try to fill this gap by studying a simple
hybrid algorithm: GES restricted to the search space determined by the
conditional independence graph.
\cite{NandyEtAl16-ARGES} show that this algorithm (and also MMHC) is \emph{not}
consistent.  Indeed, even though the global optimum lies
within the restricted search space (since the skeleton of the CPDAG is
a subgraph of the conditional independence graph), the greedy search
path may not find this global optimum without leaving the restricted
search space.  \cite{NandyEtAl16-ARGES} introduce a new hybrid
algorithm, called Adaptively Restricted GES, which was shown to be consistent in
classical and high-dimensional settings.
% \begin{marginnote}[]
% \entry{Hybrid algorithms}{Good scaling and estimation performance with theoretical guarantees}
% \end{marginnote}

\subsection{Structural Equation Models With Additional Restrictions}
\label{sec:learn-direct-lingam}

So far, we have discussed learning the Markov equivalence class of DAGs, as described by a CPDAG. Under some additional assumptions, it is possible to identify the unique DAG. To obtain intuition, we consider a simple example.

\begin{figure}
  \centering\captionsetup[subfigure]{width=1in}
    \subfloat[DAG $X \to Y$]{%
    \label{fig:lingamXtoY}%
    \includegraphics[width=4cm]{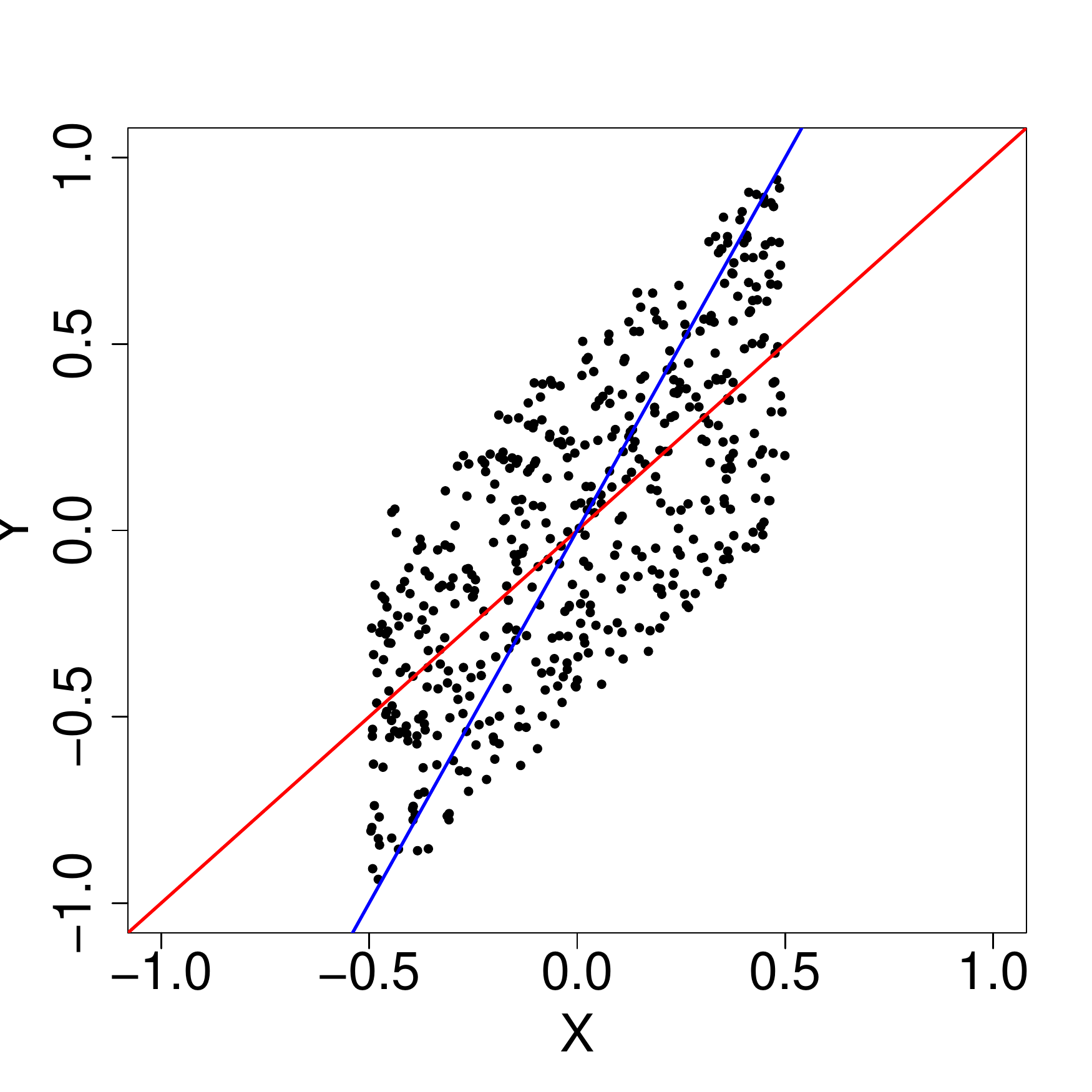}}
  \hspace{1.5cm}
    \subfloat[DAG $Y \to X$]{%
    \label{fig:lingamYtoX}%
    \includegraphics[width=4cm]{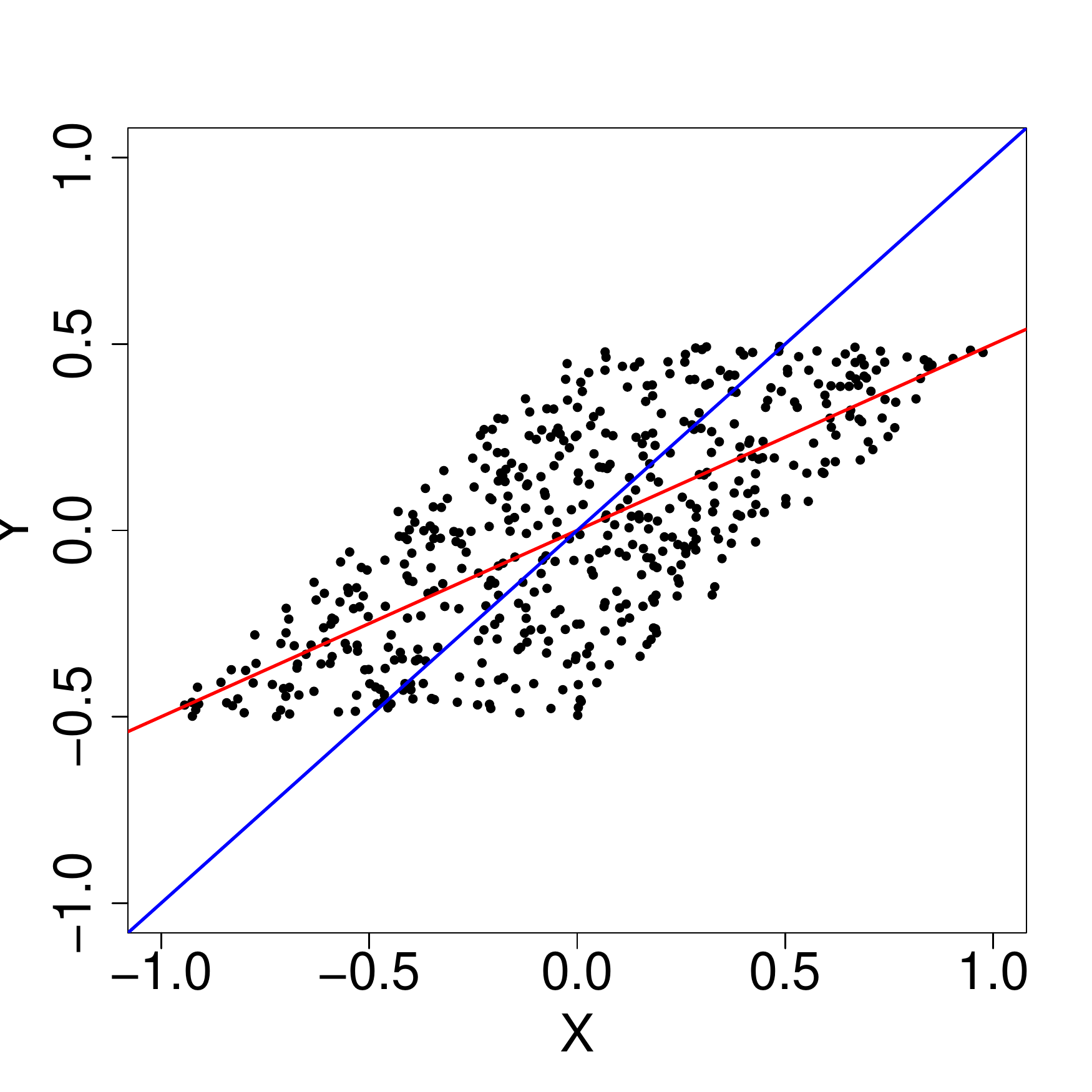}}
  \caption{Simulated data from linear structural equation models with uniform errors. Red lines indicate the regression of $Y$ on $X$, while blue lines indicate the regression of $X$ on $Y$. }
  \label{fig:lingam}
\end{figure}

\begin{example}
  Figure \ref{fig:lingamXtoY} shows a sample from the linear SEM: $X = \epsilon_X$, $Y = X + \epsilon_Y$, with $\epsilon_X$ and $\epsilon_Y$ i.i.d.\ Uniform(-0.5,0.5). The corresponding DAG is $Y \leftarrow X$, and we note that $\epsilon_Y \indep X$.
  Figure \ref{fig:lingamYtoX} shows a sample from the analogous model with $X\leftarrow Y$, where $\epsilon_X \indep Y$. The joint distributions are clearly different for the two different DAGs.
  (If the errors were Gaussian, however, the point clouds would be football shaped, and we would  not be able to distinguish the two DAGs.)

  Now suppose we are told that Figure \ref{fig:lingamXtoY} is generated from a linear SEM, and we are asked to decide whether the corresponding DAG is $X\to Y$ or $Y\to X$.
  If the DAG were $Y\to X$, then regressing $X$ on $Y$ would yield residuals that are roughly independent of $Y$. This is clearly not the case in Figure \ref{fig:lingamXtoY}. If the DAG were
  $X\to Y$, then regressing $Y$ on $X$ would yield residuals that are roughly independent of $X$. This is indeed the case in Figure \ref{fig:lingamXtoY}. Hence, we can learn that the DAG is  $X\to Y$.
\end{example}

Such ideas were first used for linear SEMs with non-Gaussian noise, or
equivalently, for Linear Non-Gaussian Acyclic Models (LiNGAM)
\citep{ShimizuEtAl06}. Recall from Example \ref{eq:SEM} that a linear
SEM can be written as $X = (I-B)^{-1}\epsilon$, meaning that $X$ is a
linear, invertible mixture of independent errors with mixing matrix
$A=(I-B)^{-1}$. In the case of non-Gaussian errors, independent
component analysis can identify $A$ up to scaling and permutation of
the columns. This idea forms the basis of the original LiNGAM
algorithm.  \cite{ShimizuEtAl11} give a more advanced implementation
that is suitable for larger numbers of variables.
The LiNGAM approach has also been extended to allow for latent
variables, time series data and feedback loops; see \cite{Shimizu14}
for an overview.
% \begin{marginnote}[]
%   \entry{LiNGAM}{For linear SEMs with additive \emph{non-Gaussian} noise, the DAG is identifiable}
% \end{marginnote}

Identifiability of the DAG can also be achieved by various other restrictions on SEMs with additive noise, such as non-linear structural equations \citep{HoyerEtAl09}, additive models \citep{BuehlmannEtAl2014}, or equal error variances \citep{PetersBuehlmann14}.

%%%%%%%%%%%%%%%%%%%%%%%%%%%%%%%%%%%

\section{Latent Variables}
\label{sec:latent-variables}

The methods discussed so far rely on data being available for all
relevant variables.  However, many applications of graphical models
involve latent, i.e., unobserved variables.  Sometimes, these are
specific variables of interest, e.g., features of extinct species in
phylogenetics.
In other settings, there may simply be a concern that observed
correlations are induced by latent variables.   We will
review some of the ideas proposed to address this latter issue.

\subsection{Low-Rank Structure in Undirected Graphical Models}
\label{sec:latent-undirected}

Let $X=(X_v:v\in V)$ be multivariate normal with covariance matrix
$\Sigma$, and let $K=\Sigma^{-1}$.  Suppose only the variables indexed
by $O\subsetneq V$ are observed.  Letting $H=V\setminus O$, we have that
$X_O=(X_v:v\in O)$  is multivariate normal with inverse covariance matrix
\begin{equation}
  \label{eq:chandra}
  \left(\Sigma_{O, O}\right)^{-1} \;=\;
  K_{O,O}-K_{O,H}\left(K_{H,H}\right)^{-1}K_{H,O}.
\end{equation}
If $X$ satisfies the Markov property with respect to an undirected graph
$G=(V,E)$, then the matrix $K_{O,O}$ is supported over the induced
subgraph $G_O$, i.e., the $(v,w)$ entry of $K_{O,O}$ is nonzero only
if $v=w$ or $v-w\in E$.  In contrast, the matrix
$K_{O,H}\left(K_{H,H}\right)^{-1}K_{H,O}$ will typically be dense as
its $(v,w)$ entry is generically nonzero whenever the graph $G$
contains a path $v-h_1-\dots-h_k-w$ with $h_1,\dots,h_k\in H$.
However, $K_{O,H}\left(K_{H,H}\right)^{-1}K_{H,O}$ has rank at most
$|H|$.  Therefore, if $G_O$ is sparse and the number of latent
variables is small, then the inverse covariance matrix of $X_O$ is the
sum of a sparse and a low-rank matrix.

% \begin{marginnote}[]
%   \entry{Latent variables in undirected Gaussian models}{Low-rank contribution
%     to inverse covariance}
% \end{marginnote}

\begin{example}
  Taking up Example~\ref{ex:ug-gaussian}, let $(X_1,\dots,X_5)$ be
  multivariate normal with the graph from
  Figure~\ref{fig:ug:exampleMP} as conditional independence graph.
  Let $K=(\kappa_{vw})$ be the inverse covariance matrix.  If
  $O=\{1,2,4,5\}$ and $H=\{3\}$, then $X_O$ has inverse covariance
  matrix
  \[
    \begin{pmatrix}
 \kappa_{11} & \kappa_{12} & 0 & 0 \\
 \kappa_{12} & \kappa_{22} & 0 & 0 \\
 0 & 0 & \kappa_{44} & \kappa_{45} \\
 0 & 0 & \kappa_{45} & \kappa_{55} \\
\end{pmatrix} -
\frac{1}{\kappa_{33}}
\begin{pmatrix}
 \kappa_{13} \\
 \kappa_{23} \\
 \kappa_{34} \\
 0 \\
    \end{pmatrix}
\begin{pmatrix}
 \kappa_{13} \\
 \kappa_{23} \\
 \kappa_{34} \\
 0 \\
\end{pmatrix}^T.
  \]
  Here, the first matrix is sparse and supported over
  the subgraph with node 3 removed, and the second matrix has rank $|H|=1$.  In
  this example, the low-rank matrix has two zero entries corresponding
  to the pairs $(X_1,X_5)$ and $(X_2,X_5)$.  This can be seen from the
  global Markov property because $X_5$ can be separated from
  $(X_1,X_2)$ by the observed variable $X_4$.
\end{example}

Suppose we have an i.i.d.~sample of size $n$ from the distribution of
$X_O$, with sample covariance matrix $S$, and wish to learn (i) the
edges between nodes $v,w\in O$ in the conditional independence graph
of $X=(X_O,X_H)$ and (ii) the number of latent variables $|H|$.
By~(\ref{eq:chandra}), both tasks can be solved simultaneously by
estimating a `sparse plus low-rank' decomposition of the inverse
covariance matrix of $X_O$.
\cite{MR3059067} propose a penalized maximum likelihood approach in
which one solves
\begin{equation}
  \label{eq:chandra-cvx}
   \min_{K^{\rm sp},K^{\rm
      lr}} \;\Big\{-\log\det(K^{\rm sp}-K^{\rm lr}) + \text{tr}\left[S(K^{\rm sp}-K^{\rm lr})\right]
+ \lambda
  \left[ \gamma\|K^{\rm sp}\|_1+\text{tr}(K^{\rm lr})\right]\Big\},
\end{equation}
subject to $K^{\rm sp}-K^{\rm lr}$ being positive definite and
$K^{\rm lr}$ being positive semidefinite. Here, $K^{\rm sp}$ and $K^{\rm lr}$ stand for the sparse and low rank components of $K$, and have corresponding $\ell_1$- and trace/nuclear norm penalties.
 For tuning parameters
$\lambda,\gamma\ge 0$, this optimization problem is convex.  Let
$(\hat K^{\rm sp},\hat K^{\rm lr})$ be a minimizer.
\cite{MR3059067} show that under
identifiability conditions the sparsity pattern of $\hat K^{\rm sp}$
and the rank of $\hat K^{\rm lr}$ consistently estimate the subgraph
$G_O$ and the number of hidden variables.  The theory covers settings
in which $|O|$ may roughly be as large as $n$.  Possible modifications
of the procedure
were proposed in discussion pieces published along with
\cite{MR3059067}.  Larger instances of~(\ref{eq:chandra-cvx}) can be
solved using an ADMM algorithm \citep{MR3100000}.
\medskip

The Gaussian example we treated is only one very special case of
graphical modeling with latent variables.  Indeed, many mixture and
latent factor models can be thought of as graphical models with latent
variables.\footnote{The states of discrete latent variables index
  different mixture components.}  Nevertheless, the example
illustrates a general phenomenon: latent variables induce low-rank
structure in tensors of moments.  This fact is exploited also in
methods of \cite{MR3270750} and \cite{chaganty2014graphical}.

\subsection{Latent Variables in Directed Graphical Models}
\label{sec:latent-directed}

If a DAG model has latent variables, then the marginal
distribution of the observed variables can generally not be
represented by a DAG. Moreover, if the marginal distribution can be
represented by a DAG, this DAG may have no causal interpretation.

\begin{example}
  \label{ex:latent-directed}
  Let the DAG $G$:
  $1 \to 2 \leftarrow 3 \to 4 \leftarrow 5$ be a perfect map of the distribution of $(X_1,\dots,X_5)$, and suppose that
  $X_3$ is latent. There is no DAG on $\{1,2,4,5\}$  that encodes
  exactly the same d-separation relations among $\{1,2,4,5\}$ as $G$. Hence, there does \emph{not} exist a
  perfect map of the marginal distribution of $(X_1,X_2,X_4,X_5)$.
\end{example}

\begin{example}
  Let the DAG $G$:
  $1 \leftarrow 2 \rightarrow 3 \leftarrow 4 \rightarrow 5$ be a perfect map of the distribution of
  $(X_1,\dots,X_5)$, and suppose that $X_2$ and $X_4$ are latent. The only conditional
  independence among the observed variables is $X_1 \indep X_5$, which
  is encoded by the DAG $G'$: $1 \to 3 \leftarrow 5$. Indeed,
  $G'$ is a perfect map of the distribution of $X=(X_1,X_3,X_5)$ and would be found when
  applying consistent methods such as the PC algorithm to a large
  sample of $(X_1,X_3,X_5)$.  However, $G'$ does \emph{not} reflect the causal
  interpretation of $G$.  For example, $G'$ suggests
  $X_1$ as a cause of $X_3$, but there is no directed path from $X_1$
  to $X_3$ in $G$.
\end{example}

Mixed graphs provide a useful approach to address these problems
without explicit modeling of latent variables
\citep[e.g.,][]{Pearl09,SpirtesEtAl00,MR2817608}.  The nodes of these
graphs index the observed variables only.  The edges, however, may be
of two types, directed and bidirected.  This added flexibility allows
one to represent the more complicated dependence structures arising
from a DAG with latent variables.  A straightforward generalization of
d-separation determines conditional independencies in mixed graph
models.  For instance, the mixed graph
$1\to 2 \longleftrightarrow 4\leftarrow 5$ is a perfect map for the distribution in
Example~\ref{ex:latent-directed}.

To facilitate constraint-based structure learning in settings with
latent variables, \cite{RichardsonSpirtes02}
introduce a class of mixed graphs known as \emph{maximal ancestral
  graphs} (MAGs).\footnote{The work also considers selection bias
  which we here ignore.}  Every DAG $G=(V,E)$ with $V= O \cup H$,
where $O$ and $H$ index the observed and latent variables,
respectively, can be transformed into a unique MAG on $O$ such that
conditional independencies among the observed variables are
preserved.
The MAG also encodes ancestral relationships in the underlying DAG
$G$, as follows.  If the MAG has edge $v\to w$, then $v \in \an_G(w)$ but
$w \notin \an_G(v)$. Similarly, $v\leftrightarrow w$
encodes $v\notin \an_G(w)$ and $w\notin \an_G(v)$.  In general,
several MAGs may describe the same set of conditional independencies.
The resulting Markov equivalence class of MAGs can be described by a
Partial Ancestral Graph (PAG) \citep{AliEtAl09}.

PAGs can be learned by a generalization of the PC algorithm, called
the FCI algorithm \citep{SpirtesEtAl99,RichardsonSpirtes02}.  As noted
in Section~\ref{sec:learn-direct-constraint-based}, the PC algorithm
is of polynomial time for graphs of bounded degree, by exploiting the
fact that the edge $v-w$ is absent in the skeleton of the DAG $G$ if
and only if $X_v \indep X_w \,|\, X_{\pa_G(v)}$ or
$X_v \indep X_w \,|\, X_{\pa_G(w)}$. In the presence of latent
variables, this fact no longer holds. The FCI algorithm therefore
performs additional tests, and the number of such tests can be
exponential in the number of nodes, even for sparse graphs. The FCI
algorithm also uses more complicated orientation rules, which were
extended and proved to be complete by
\cite{Zhang08-orientation-rules}. \cite{ColomboEtAl12} and \cite{ClaassenEtAl13} introduce
fast modifications of the FCI algorithm that are of polynomial time for graphs of bounded degree.
% \begin{marginnote}[]
%   \entry{DAGs with latent variables}{Mixed graphs represent observable
%     conditional independencies}
%   \entry{FCI algorithm}{Learns an equivalence class of mixed graphs}
% \end{marginnote}

While MAGs can represent all conditional independencies in the
marginal distribution of the observed variables, there can be equality
and inequality constraints that MAGs cannot represent. An example is
the so-called Verma constraint
\citep{VermaPearl90}; see also
Example 3.3.14 in \cite{MR2723140}.  To represent constraints beyond
conditional independencies, there is current work on new classes of
graphs, including nested Markov models
\citep{ShpitserEtAl12,EvansEtAl14} and mDAGs \citep{Evans16}.

%%%%%%%%%%%%%%%%%%%%%%%%%%%%%%%%%%%

\section{Heterogeneous Data}
\label{sec:heterogeneous-data}

Heterogeneous data that do not form an i.i.d.~sample from a single
population are encountered, for instance, in gene expression studies
involving different organisms or experimental conditions, or in the
comparative analysis of brain networks for patients with different
neurological disorders.  In such settings, it is of interest to learn
the structure of graphical models for subpopulations.  More
generally, a graph may depend on covariates.

\cite{MR2804206} propose an extension of the graphical lasso
from~(\ref{eq:glasso}) to estimate undirected conditional independence
graphs of several related Gaussian populations.  The authors sum up
the log-likelihood functions for $m$ populations with inverse
covariance matrices $K_1,\dots,K_m\in\mathbb{R}^{V\times V}$ and then
add a sparsity-inducing penalty.  To share common structure, the
inverse covariances are reparametrized as
$K_{i,vw}=\theta_{vw}\gamma_{i,vw}$.  The penalty then adds the
(vector) $\ell_1$ norms of the matrix $(\theta_{vw})$ and the $m$
matrices $(\gamma_{i,vw})$.  Sparsity in an estimate of
$(\theta_{vw})$ results in edges being simultaneously absent from all
$m$ graph estimates, each of which may have further edges absent
through zero estimates of $\gamma_{i,vw}$.  A downside of this method
is that its optimization problem is not convex.  \cite{MR3164871}
propose instead the use of group or fused lasso penalties specified in
terms of the inverse covariances.
The group lasso
penalty takes the form
$ \sum_{v\not=w} \sqrt{K_{1,vw}^2+\dots+K_{m,vw}^2} $ and leads to
edges being simultaneously absent from all $m$ graph
estimates;  a version of this approach that separates positive from
negative signals was also proposed by \cite{MR2826691}.  The
fused lasso penalty
$ \sum_{1\le i<j\le m}\sum_{v,w} |K_{i,vw}-K_{j,vw}| $ yields pairwise
similar edge patterns in the different populations.  \cite{MR3343365}
consider generalizations of this fused lasso penalty and characterize
when the resulting optimization problem can be decomposed into smaller
problems arising from block-diagonal inverse covariance matrices.
\cite{Saegusa:2016} treat settings in
which some populations may be more closely related than others.
Finally, \cite{MR3215346} show how the difference between two conditional
independence graphs can be estimated without estimating the two
graphs.

% \begin{marginnote}[]
%    \entry{Heterogeneous data}{Grouped samples, time-varying graphs, different experimental conditions}
% \end{marginnote}

Heterogeneous data also arise from time-course observations, for which
we may wish to learn time-varying structure.
\cite{MR3108169} compute glasso estimates at different time points,
taking as input a weighted sample covariance matrix that is a
kernel estimate of the covariance matrix at time $t$.  Alternatively,
\cite{MR2758086} and \cite{MR3020257} use fused lasso penalties.
Matrix/tensor-normal models, in which one of
the dimensions could be time, constitute another approach; see
\cite{MR3199836} and references therein.

Similar issues arise for directed graphical models.  In particular,
so-called dynamic Bayesian networks have long been used to model
temporal dependencies, and there is a natural connection to
vector-autoregressive processes in the time series literature
\citep[e.g.,][]{ShojaieGranger}.  A detailed discussion %of this topic
is beyond the scope of this paper.

Finally, we
note that heterogeneity from different experimental conditions can
also be exploited in structure learning for directed graphical models \citep{DanksGlymourTillman09, HauserBuehlmann12, HyttinenEtAl12, TriantafillouTsamardinos15}. These papers generally assume to have i.i.d.\ observations from various known experimental conditions, and some of them allow cycles and/or latent variables.
\cite{Peters} study a scenario where data may come from different experimental conditions, but these conditions are unknown. They provide a method built on the idea that a
variable can be well predicted from its causes across different
experimental conditions.

%%%%%%%%%%%%%%%%%%%%%%%%%%%%%%%%%%%

\section{Discussion}
\label{sec:discussion}

Stimulated to a large extent by applications in gene expression
analysis, the field of structure learning in graphical modeling has
undergone rapid development, with much of the new work focusing on
high-dimensional problems.  In this review, we treated some of
the main ideas behind these developments, including
$\ell_1$-regularization techniques, greedy search approaches and
methods based on conditional independence tests.  Extensions to cope
with latent variables have long been of interest and continue to be
addressed in new ways.  More recently, challenges arising in
connection with heterogeneous and dependent data have inspired methods
that generalize those for i.i.d.~data.  Similarly, methods for
continuous but non-Gaussian data are an active area of research.

In a different vein, it is of interest to provide an uncertainty
assessment for estimates of graph structure.  Bayesian approaches
naturally include an uncertainty assessment but frequentist techniques
that are able to cope with high-dimensional data are also being
developed \citep{MR3354336,MR3346695,MR3371002}.

Finally, our treatment of directed graphical models considered acyclic
graphs, in which no feedback loops exist in the cause-effect
relationships that the model captures.  Effective methods for
structure learning exist for the acyclic case, but coping with
feedback loops is a far more difficult problem.  While certain forms
of feedback can be represented in the paradigm of linear structural
equation modeling \citep{SpirtesEtAl00,MooijHeskes_UAI_13} and
conditional independence can then be exploited in structure learning
\citep{Richardson96}, these models can generally not be described using
solely conditional independence; see Example 3.6 and Appendix A in
\cite{MR2502658}.  New ideas are still needed to effectively learn
cyclic cause-effect relationships from possibly high-dimensional
observational data.

% \begin{issues}[FUTURE ISSUES]
% \begin{enumerate}
% \item Methods that account for the effects of latent variables need to
%   be developed further.
% \item Emerging techniques for formal statistical inference for
%   high-dimensional graphical models should see continued
%   development.
% \item New methods for cyclic directed graphical models are needed to
%    effectively learn cause-effect networks with feedback loops.
% \end{enumerate}
% \end{issues}

\small
\bibliographystyle{ar-style1}
\bibliography{mathias,marloes}

\end{document}

%% file: macros.tex
\DeclareMathOperator{\E}{\mathbb{E}}

\newtheorem{theorem}{Theorem}[section]

\theoremstyle{definition}
\newtheorem{definition}{Definition}[section]

\theoremstyle{remark}

\newtheorem{example}{Example}[section]

\makeatletter
\newtheorem*{rep@theorem}{\rep@title}
\newcommand{\newreptheorem}[2]{%
\newenvironment{rep#1}[1]{%
 \def\rep@title{#2 \ref{##1}}%
 \begin{rep@theorem}}%
 {\end{rep@theorem}}}
\makeatother

\newreptheorem{theorem}{Theorem}
\newreptheorem{lemma}{Lemma}
\newreptheorem{corollary}{Corollary}

\newcommand\indep{\protect\mathpalette{\protect\independenT}{\perp}}
\def\independenT#1#2{\mathrel{\rlap{$#1#2$}\mkern3mu{#1#2}}}

\DeclareMathOperator{\cov}{Cov}

\DeclareMathOperator{\nd}{nd}
\DeclareMathOperator{\an}{an}
\DeclareMathOperator{\de}{de}
\DeclareMathOperator{\adj}{adj}
\DeclareMathOperator{\nb}{nb}
\DeclareMathOperator{\pa}{pa}

\def\dsep{\perp}